\documentclass[aps, prb, twocolumn, footinbib, superscriptaddress]{revtex4-2}

\usepackage[T1]{fontenc}
\usepackage{amsmath,amsfonts,amssymb,amsthm,bbm,bm,braket}
\usepackage{wasysym}
\usepackage{enumerate}
\usepackage{graphicx}
\usepackage{mathtools}
\usepackage{color}
\usepackage{phfqit}
\usepackage[dvipsnames]{xcolor}
\usepackage{tikz}
\usetikzlibrary{shadings, decorations.pathreplacing, arrows.meta, decorations.text, patterns}
\usepackage[colorlinks,bookmarks=false,citecolor=NavyBlue,linkcolor=OliveGreen,urlcolor=blue]{hyperref}

\definecolor{blue}{rgb}{0,0,1}
\definecolor{grey}{rgb}{0.6,0.6,0.6}
\definecolor{spinRed}{RGB}{227, 26, 28}
\definecolor{spinBlue}{RGB}{31, 120, 180}
\definecolor{rotGreen}{RGB}{0, 0, 0}
\definecolor{scLead}{RGB}{166, 206, 227}
\definecolor{scLeadDark}{RGB}{128, 177, 204}
\definecolor{geGray}{rgb}{0.72,0.71,0.71}
\definecolor{geGrayDark}{rgb}{0.39,0.38,0.38}
\definecolor{gap1}{RGB}{227, 149, 23}
\definecolor{gap2}{RGB}{20, 186, 219}

\definecolor{strongGap1}{HTML}{e35a53}
\definecolor{strongGap2}{HTML}{25b3d6}

\newcommand{\ua}{\uparrow}
\newcommand{\da}{\downarrow}

\newcommand\dif[1]{\mathrm{d}#1\,}
\begin{document}

\title{Spin-resolved Andreev spectra in spin-orbit-coupled Josephson junctions\\ with mixed-parity pairing}

\author{Markus Plautz}
\email{markus.plautz@ntnu.no}
\affiliation{Center for Quantum Spintronics, Department of Physics, Norwegian \\ University of Science and Technology, NO-7491 Trondheim, Norway}
\author{Florinda Viñas Boström}
\author{Antonio L. R. Manesco}
\author{Karsten Flensberg}
\affiliation{Niels Bohr Institute, University of Copenhagen, DK-2100
Copenhagen, Denmark}
\author{Jeroen Danon}
\affiliation{Center for Quantum Spintronics, Department of Physics, Norwegian \\ University of Science and Technology, NO-7491 Trondheim, Norway}

\date{\today}

\begin{abstract}
Proximitized semiconductor Josephson junctions can host mixed singlet and triplet pairing, which modifies the spin-resolved spectrum of their Andreev bound states.
We develop an analytical description of quasi-one-dimensional junctions with intrinsic spin-orbit coupling and effective ($s+p$)-wave pairing, such as expected in hole-gas-based junctions. 
We derive the Andreev reflection matrices and bound-state conditions in the limits of strong and weak spin-orbit coupling, retaining the band-dependent Fermi momenta and velocities. 
In the strong-coupling regime, interband pairing is suppressed and the two spin-orbit bands acquire distinct gaps determined by the projection of 
the triplet pairing onto the spin-orbit axis. 
In the weak-coupling regime, misalignment of these axes produces spin-mixing Andreev reflection. We calculate the reflection matrices through second order in spin-orbit coupling and 
identify how the interface spin rotations combine with propagation through the normal region to determine the level splitting. 
Comparisons with real-space tight-binding calculations validate the analytical spectra and their dependence on spin-orbit strength.
Our results clarify how this mixed pairing shapes the Andreev spectrum and provide a basis for assessing its impact on devices such as Andreev spin qubits.
\end{abstract}

\maketitle

\section{Introduction}\label{sec:introduction}
Hybrid semiconductor-superconductor Josephson junctions have emerged as a versatile
platform for quantum technologies~\cite{dohTunableSupercurrentSemiconductor2005,
larsenSemiconductorNanowireBasedSuperconductingQubit2015,
delangeRealizationMicrowaveQuantum2015,
casparisSuperconductingGatemonQubit2018,
haysCoherentManipulationAndreev2021,
burkardSuperconductorSemiconductorHybridcircuit2020,
aguadoPerspectiveSemiconductorbasedSuperconducting2020,
flensbergEngineeredPlatformsTopological2021,
seoanesoutoSubgapStatesSemiconductorSuperconductor2024,
pita-vidalNovelQubitsHybrid2025}. Their subgap electronic spectrum contains Andreev bound states, 
discrete phase-dependent excitations localized in the weak link~\cite{beenakkerJosephsonCurrentSuperconducting1991,
furusakiCurrentcarryingStatesJosephson1991,saulsAndreevBoundStates2018}. The energies, spin structure, and
occupation of these states govern the low-energy response of the junction and
provide the basic degrees of freedom for several proposed quantum devices.

In quasi-one-dimensional junctions, the interplay of spin-orbit coupling (SOC), transverse confinement 
and subband mixing can give the two spin-split branches of the lowest occupied transverse 
mode different Fermi velocities~\cite{parkAndreevSpinQubits2017,
yokoyamaAnomalousJosephsonEffect2014,tosiSpinOrbitSplittingAndreev2019}.
In a finite-length junction, their
different propagation times produce a spin splitting of the Andreev spectrum at
finite phase difference even without a magnetic field~\cite{beriSplittingAndreevLevels2008,
parkAndreevSpinQubits2017,vanheckZeemanSpinorbitEffects2017,
vanwoerkomMicrowaveSpectroscopySpinful2017,tosiSpinOrbitSplittingAndreev2019}. Spin-resolved
Andreev spectra probe the spin-orbit-coupled band
structure~\cite{vanwoerkomMicrowaveSpectroscopySpinful2017,
tosiSpinOrbitSplittingAndreev2019}, reveal parity-changing
transitions~\cite{leeSpinresolvedAndreevLevels2014a} and protected level
crossings~\cite{muraniAndreevSpectrumHigh2017}, while also influencing the
anomalous~\cite{reynosoSpinorbitinducedChiralityAndreev2012,
yokoyamaAnomalousJosephsonEffect2014,szombatiJosephsonPh0junctionNanowire2016} and nonreciprocal
Josephson effects~\cite{costaSignReversalJosephson2023,
reinhardtLinkSupercurrentDiode2024,patilSpinsplitAndreevBound2025}. They also form the basis of Andreev spin
qubits~\cite{chtchelkatchevAndreevQuantumDots2003,
padurariuTheoreticalProposalSuperconducting2010,parkAndreevSpinQubits2017}, for
which microwave spectroscopy and coherent spin control have been demonstrated
experimentally~\cite{vanwoerkomMicrowaveSpectroscopySpinful2017,
tosiSpinOrbitSplittingAndreev2019,
haysCoherentManipulationAndreev2021,pita-vidalDirectManipulationSuperconducting2023}.

Microscopic studies have, however, shown that SOC can generate
proximity-induced triplet correlations and momentum-dependent pairing components
in semiconductors, such that the induced pairing need not remain purely conventional
$s$-wave~\cite{reegProximityinducedTripletSuperconductivity2015,
yuGappedTriplet$p$wave2016,phanDetectingInduced$pifmmodepmelsetextpmfiip$2022,
mishraEffectsSpinOrbit2023,babkinSuperconductingProximityEffect2025,
pinoTheorySuperconductingProximity2025}. Mixed singlet-triplet pairing and its
effect on Andreev states have been studied in noncentrosymmetric
superconductors~\cite{frigeriSuperconductivityInversionSymmetry2004,
iniotakisAndreevBoundStates2007,asanoJosephsonEffectNoncentrosymmetric2011,
klamJosephsonEffectTripletsinglet2014,smidmanSuperconductivitySpinOrbit2017},
in explicit $(s+p)$-wave junction models~\cite{bursetTransportSignaturesSuperconducting2014,
guanBandlikeExactZeroenergy2025}, and in proximitized topological-insulator
junctions~\cite{fuSuperconductingProximityEffect2008,
tkachovHelicalAndreevBound2013}. Its effect on the spin-resolved
Andreev spectrum of semiconductor weak links with intrinsic SOC
remains less well understood.

Here we develop an analytical description of a ballistic one-dimensional
semiconductor Josephson junction with mixed $(s+p)$-wave pairing and SOC that can include terms both linear and cubic in momentum. We derive the
Andreev reflection matrices, bound-state conditions and level splitting in the
limits of strong and weak spin-orbit coupling. These limits are set by the ratio of the spin-orbit energy, which sets the separation of the two spin-orbit bands at the Fermi level, to the induced gap. 
In the strong-SOC regime, the
triplet component gives the two spin-orbit bands different gaps. The resulting splitting
persists in the short-junction limit, where the splitting from a velocity mismatch vanishes.
In the weak-SOC regime, a misalignment of the spin-orbit and triplet-pairing axes instead
generates spin-mixing Andreev reflection. We show how triplet pairing and the different propagation
times in the two bands jointly determine the level splitting and its dependence on the SOC strength.
We benchmark these results against real-space tight-binding calculations.

Finally, we apply our results to germanium hole wires. Effective $(s+p)$ pairing has recently been predicted for
proximitized germanium hole gases~\cite{babkinSuperconductingProximityEffect2025,pinoTheorySuperconductingProximity2025},
which have also emerged as a promising platform for superconductor-semiconductor
devices~\cite{scappucciGermaniumQuantumInformation2021,aggarwalEnhancementProximityInduced2021,
tosatoHardSuperconductingGap2023,hinderlingDirectMicrowaveSpectroscopy2024,lakicQuantumDotGermanium2025,
hoffmanResolvingAndreevSpin2025,fabrisGranularAluminumInduced2026, vivrekarGateTunableSpin2026}.
We extract the effective parameters from a microscopic Luttinger-Kohn model and find that such wires lie deep in the
strong-SOC regime. Both the triplet pairing and the velocity mismatch contribute to the level splitting. Their relative weight depends on details of the device such as strain.

The remainder of this paper is organized as follows. Section~\ref{sec:model}
introduces the effective model. Section~\ref{sc:results} derives the Andreev
spectrum in the strong- and weak-SOC regimes and compares the analytical results
with tight-binding calculations. Section~\ref{sec:ge} connects the effective
parameters to a microscopic model of a proximitized germanium hole gas. We
summarize our findings in Sec.~\ref{sc:conclusion}.

\section{Model}\label{sec:model}

\begin{figure}[!t]
    \centering
    \begin{tikzpicture}[scale=0.95, every node/.style={font=\small}]
        \def\xl{-3.2}  \def\xr{3.2}      
        \def\xa{-1.2}  \def\xb{1.2}      
        \def\hg{0.32}                    
        \def\hs{0.38}                    
        \def\dx{0.5}   \def\dy{0.3}      

        \node at (\xl-0.9, 0.95) {(a)};

        \fill[geGray] (\xl,0) -- (\xr,0) -- ++(\dx,\dy) -- (\xl+\dx,\dy) -- cycle;
        \fill[geGray] (\xl,-\hg) rectangle (\xr,0);
        \fill[geGray] (\xr,-\hg) -- (\xr,0) -- ++(\dx,\dy) -- ++(0,-\hg) -- cycle;
        \fill[pattern=north east lines, pattern color=geGrayDark] (\xr,-\hg) -- (\xr,0) -- ++(\dx,\dy) -- ++(0,-\hg) -- cycle;
        \fill[pattern=north east lines, pattern color=geGrayDark] (\xl,-\hg) rectangle (\xa,0);
        \fill[pattern=north east lines, pattern color=geGrayDark] (\xb,-\hg) rectangle (\xr,0);
        \draw[thick, geGrayDark] (\xa,0) -- ++(0,-\hg+0.01);
        \draw[thick, geGrayDark] (\xb,0) -- ++(0,-\hg);
        \draw[thick] (\xl,-\hg) rectangle (\xr,0);
        \draw[thick] (\xr,-\hg-0.005) -- ++(\dx,\dy) -- ++(0,\hg);
        \draw[thick] (\xa,\dy) ++(\dx-0.5,0) -- (\xb+\dx,\dy); 
        \node at (0,-0.5*\hg) {N};

        \fill[scLead] (\xl,0) rectangle (\xa,\hs);
        \fill[scLead] (\xl,\hs) -- (\xa,\hs) -- ++(\dx,\dy) -- (\xl+\dx,\hs+\dy) -- cycle;
        \fill[scLeadDark] (\xa,0) -- (\xa,\hs) -- ++(\dx,\dy) -- ++(0,-\hs) -- cycle;
        \draw[thick] (\xl,0) rectangle (\xa,\hs);
        \draw[thick] (\xl,\hs) -- ++(\dx,\dy) -- (\xa+\dx-0.025,\hs+\dy-0.01) -- (\xa,\hs);
        \draw[thick] (\xa+\dx,\hs+\dy) -- ++(0,-\hs);
        \draw[thick] (\xa, 0) -- ++(\dx,\dy);
        \node at (0.5*\xl+0.5*\xa,0.5*\hs) {$\varphi_{\mathrm{L}}$};

        \fill[scLead] (\xb,0) rectangle (\xr,\hs);
        \fill[scLead] (\xb,\hs) -- (\xr,\hs) -- ++(\dx,\dy) -- (\xb+\dx,\hs+\dy) -- cycle;
        \fill[scLeadDark] (\xr,0) -- (\xr,\hs) -- ++(\dx,\dy) -- ++(0,-\hs) -- cycle;
        \draw[thick] (\xb,0) rectangle (\xr,\hs);
        \draw[thick] (\xb,\hs) -- ++(\dx,\dy) -- (\xr+\dx-0.03,\hs+\dy) -- (\xr,\hs);
        \draw[thick] (\xr+\dx,\hs+\dy) -- ++(0,-\hs);
        \draw[thick] (\xr,0) -- ++(\dx,\dy);
        \node at (0.5*\xb+0.5*\xr,0.5*\hs) {$\varphi_{\mathrm{R}}$};

        \begin{scope}[shift={(0,0.5*\dy)}]
            \draw[->, >={Triangle[scale=0.6]}] (0,0) -- (-0.35,0.62) node[above] {$\hat{\boldsymbol{B}}_{\mathrm{Z}}$};
            \draw[->, >={Triangle[scale=0.6]}] (0,0) -- (0.35,0.62) node[above] {$\hat{\boldsymbol{m}}_{\mathrm{so}}$};
            \filldraw[black] (0,0) circle (1.3pt);
        \end{scope}

        \draw[<->, >={Triangle[scale=0.7]}, line width=0.5pt] (\xa,-\hg - 0.15) -- ++(\xb - \xa, 0);
        \node[below] at (\xa + \xb*0.5 - \xa*0.5, -\hg-0.15) {\footnotesize $L$};

        \begin{scope}[shift={(\xl-1.05,-0.3)}, >={Triangle[scale=0.5]}]
            \draw[->] (0,0) -- (0.5,0) node[below] {\footnotesize $x$};
            \draw[->] (0,0) -- (0.4,0.24) node[above right=-2pt] {\footnotesize $y$};
            \draw[->] (0,0) -- (0,0.5) node[above] {\footnotesize $z$};
        \end{scope}
    \end{tikzpicture}

    \vspace{0.1cm}

    \includegraphics[width=1\linewidth,height=108pt]{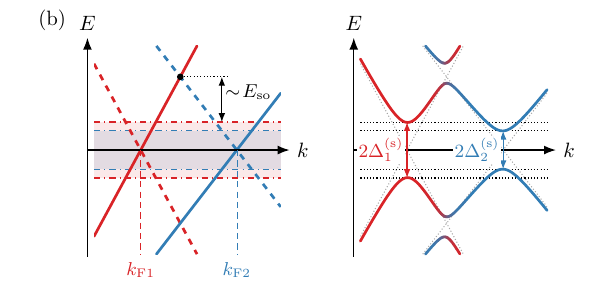}

    \vspace{0.1cm}

    \includegraphics[width=1\linewidth,height=108pt]{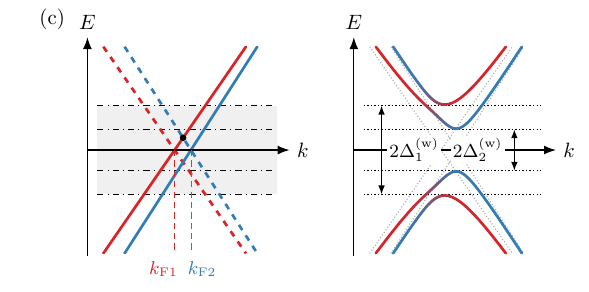}
    \caption{(a) Sketch of the setup considered. A semiconducting wire with a rectangular
    cross-section is covered at both ends by superconducting leads with phases~$\varphi_{\mathrm{L}}$
    and~$\varphi_{\mathrm{R}}$, which proximitize the wire underneath (hatched, S). The uncovered normal region (N) of
    length~$L$ forms the weak link. The wire is subject to a linear and cubic spin-orbit field, pointing along a common
    axis~$\hat{\boldsymbol{m}}_{\mathrm{so}}$, 
    and to a Zeeman field pointing along $\hat{\bf B}_{\rm Z}$ acting only in the normal region.
    (b)~Strong spin-orbit coupling. The left panel shows the electron (solid) and hole (dashed) branches of the two
    bands near their Fermi points at positive $k$. Shaded areas mark the ``pairing windows''~$\pm\Delta_{1,2}^{(\mathrm{s})}$ of Eq.~\eqref{eq:helicity_gaps}. An electron branch of one band
    crosses a hole branch of the other (black dot) at a distance of order~$E_{\mathrm{so}}$ above the gap edge. The
    right panel shows the resulting BdG spectrum in the superconductor, colored by band weight. Each band is gapped
    by its own~$\Delta_j$. The interband pairing~$\Delta_{12}$ only opens anticrossings at the distant crossings and
    may thus be neglected for~$\abs{\Delta_{12}}\ll E_{\mathrm{so}}$.
    (c)~Weak spin-orbit coupling. The interband crossing now lies within the pairing windows. The pairing mixes the
    two bands, such that the gaps are given by the eigenvalues~$\Delta_{1,2}^{(\mathrm{w})}=\Delta_{\mathrm S}\pm\Delta_{\mathrm P}$
    of the pairing matrix. Gray dotted lines in the right panels show the bare branches.}
    \label{fig:setup}
\end{figure}

The setup we consider is a one-dimensional semiconducting nanowire connected to two superconducting contacts, as depicted in
Fig.~\ref{fig:setup}(a).
The leads, located at $x<0$ and $x>L$, induce a superconducting pairing potential in the nanowire via the proximity effect.
We take the induced gaps to have equal magnitude on both sides of the junction and choose the chemical potential such that only 
the lowest transverse mode is occupied. For brevity, we refer to the two pseudospin states within this mode as spin up and spin down 
in the following.

We describe our system using a Bogoliubov-de Gennes (BdG) Hamiltonian of the
form~$H=\frac{1}{2}\int \dif{x}\boldsymbol{\Psi}^{\dagger} \hat{\mathcal{H}}_{\mathrm{BdG}}\boldsymbol{\Psi}$,
where~$\boldsymbol{\Psi}=[\psi_{\uparrow}(x),\psi_{\downarrow}(x), \psi_{\downarrow}^{\dagger}(x), -\psi_{\uparrow}^{\dagger}(x)]^{\top}$
with~$\psi_{\uparrow(\downarrow)}(x)$ being the annihilation operator for a charge carrier at position~$x$ with spin up (down).
With this, we have
\begin{equation}\label{eq:BdG}
    \hat{\mathcal{H}}_{\mathrm{BdG}}(x) =
    \begin{pmatrix}
        \hat{H}_0 & \hat{\Delta}(x) \\
        \hat{\Delta}^\dagger(x) & -\mathcal{T}\hat{H}_0\mathcal{T}^{\dagger}
    \end{pmatrix},
\end{equation}
where $\hat{H}_0(k)$ is the single-particle Hamiltonian, $\hat{\Delta}(x)$ the pairing potential
and~$\mathcal{T}$ the time-reversal operator.
Throughout, $k=-i\partial_x$ denotes the momentum operator.

We use the effective one-dimensional Hamiltonian
\begin{equation}\label{eq:H0_full}
    \hat{H}_0 = \frac{\hbar^2 k^2}{2m} - \mu
    + (k\alpha_{\mathrm{so}} + k^3\beta_{\mathrm{so}})\,\hat{\boldsymbol{m}}_{\mathrm{so}}\cdot\boldsymbol{\sigma}
    + \mathbf{B}_{\mathrm{Z}}\cdot\boldsymbol{\sigma},
\end{equation}
with $\boldsymbol{\sigma}$ the vector of Pauli matrices
in spin space.
The parameters~$\alpha_{\mathrm{so}}$ and~$\beta_{\mathrm{so}}$ characterize the linear and cubic spin-orbit coupling, which are the two lowest-order contributions allowed by time-reversal symmetry.
The linear Rashba term stems from structural inversion asymmetry, for example a gate field perpendicular to the wire~\cite{kloeffelStrongSpinorbitInteraction2011,kloeffelDirectRashbaSpinorbit2018}.
The cubic term can arise from spin-orbit-induced mixing with excited states, such as higher transverse subbands~\cite{parkAndreevSpinQubits2017} or split-off
light-hole states in hole-based systems~\cite{terrazosTheoryHolespinQubits2021}. 
We take both terms to share a common axis~$\hat{\boldsymbol{m}}_{\mathrm{so}}$, as is the case when a single source of inversion asymmetry, such as the gate field, dominates.
The linear term shifts the two spin-split bands in momentum, such that right (left) movers have the Fermi wave vectors~$\pm k_{\mathrm{F}1,2}$, and the cubic term also gives the two bands different Fermi velocities~$v_{\mathrm{F}1,2}$.
The vector~$\mathbf{B}_{\mathrm{Z}}$ denotes the Zeeman field, which we take to act only in the normal region,~$0<x<L$.

We allow the pairing in the superconducting regions to contain both singlet
and triplet components. The pairing potential is thus
\begin{equation}\label{eq:pairing}
    \hat{\Delta}(x)=
    \begin{cases}
        \hat{\Delta}_{\mathrm{L}}, & x\leq 0,\\[2pt]
        0, & 0<x<L,\\[2pt]
        \hat{\Delta}_{\mathrm{R}}, & x\geq L,
    \end{cases}
\end{equation}
where
\begin{equation}\label{eq:pairing_leads}
    \hat{\Delta}_{\nu}
    =
    e^{i\varphi_{\nu}}
    \left[
        \Delta_{\mathrm{S}}\mathbbm{1}_2
        +\Delta_{\mathrm{P}}\frac{k}{k_{\mathrm{F}}}
        \left(\hat{\boldsymbol{d}}\cdot\boldsymbol{\sigma}\right)
    \right],
    \quad \nu=\mathrm{L},\mathrm{R}.
\end{equation}
Here, $\varphi_{\mathrm{L,R}}$ are the superconducting phases in the two proximitized regions, and the real amplitudes~$\Delta_{\mathrm{S}}$ and~$\Delta_{\mathrm{P}}$ and the real unit
vector~$\hat{\boldsymbol{d}}$ describe the singlet and triplet pairing. Both components are thus assumed to share the
phase~$\varphi_\nu$, and the pairing is time-reversal symmetric.
We use the Fermi wavevector
$k_{\mathrm{F}}=\sqrt{2m\mu}/\hbar$ of the underlying spin-independent parabolic dispersion as the reference momentum,
where~$m$ is the effective mass and~$\mu$ is the chemical potential.
Within either uniform superconducting region, the $k$-space pairing matrix has the 
eigenvalues~$\Delta_{1,2}(k)=\Delta_{\mathrm{S}}\pm \Delta_{\mathrm{P}}k/k_{\mathrm{F}}$.
Throughout this work, we restrict to the fully gapped regime with~$\abs{\Delta_{\mathrm{P}}}\lesssim\Delta_{\mathrm{S}}$ 
in which the pairing eigenvalues at the relevant Fermi
points are nonzero and have the same sign. 
For~$\abs{\Delta_{\mathrm{P}}}\gtrsim\Delta_{\mathrm{S}}$, the two eigenvalues have opposite signs. In minimal mixed
$(s+p)$-wave models, this sign reversal marks a transition into a topologically nontrivial, triplet-dominated regime
with zero-energy boundary and Andreev states~\cite{asanoJosephsonEffectNoncentrosymmetric2011,
bursetTransportSignaturesSuperconducting2014,guanBandlikeExactZeroenergy2025}.

\section{Results}\label{sc:results}

\subsection{Scattering approach}\label{sec:scattering}

We focus on the subgap physics of the junction and assume~$\Delta_{\mathrm{S}},\Delta_{\mathrm{P}}\ll\mu$,
such that only states in the immediate vicinity of the Fermi points contribute. This justifies the Andreev
approximation, in which we linearize Eq.~\eqref{eq:H0_full} around the reference momenta~$\pm k_{\mathrm F}$.
The Zeeman field acts only in the normal region. It therefore leaves the Andreev reflection at the interfaces
unchanged and enters only the propagation between them. We first linearize the field-free Hamiltonian, which
describes the leads and, for~$\mathbf B_{\mathrm Z}=0$, also the normal region. Close to~$+k_{\mathrm F}$ this yields
\begin{equation}\label{eq:H0_lin}
    \hat H_0\simeq\hbar\left(v_0+\delta v_{\mathrm F}\,\hat{\boldsymbol m}_{\mathrm{so}}\cdot\boldsymbol\sigma\right)(k-k_{\mathrm F})
    +E_{\mathrm{so}}\,\hat{\boldsymbol m}_{\mathrm{so}}\cdot\boldsymbol\sigma,
\end{equation}
with~$v_0=(v_{\mathrm F1}+v_{\mathrm F2})/2$ and~$\delta v_{\mathrm F}=(v_{\mathrm F1}-v_{\mathrm F2})/2$. Close
to~$-k_{\mathrm F}$ the expansion follows from time reversal, $k_{{\rm F}}\to-k_{{\rm F}}$
and~$\hat{\boldsymbol m}_{\mathrm{so}}\cdot\boldsymbol\sigma\to-\hat{\boldsymbol m}_{\mathrm{so}}\cdot\boldsymbol\sigma$,
so that a band keeps its velocity~$v_{\mathrm Fj}$ but reverses its spin. The spin-orbit
energy~$E_{\mathrm{so}}=\alpha_{\mathrm{so}}k_{\mathrm F}+\beta_{\mathrm{so}}k_{\mathrm F}^3$ is the spin-orbit field
at~$k_{\mathrm F}$. It sets the energy separation of the two bands at the Fermi level, whereas~$\delta v_{\mathrm F}$
sets their velocity difference.

\begin{figure*}[t]
    \centering
    \includegraphics[width=0.95\textwidth]{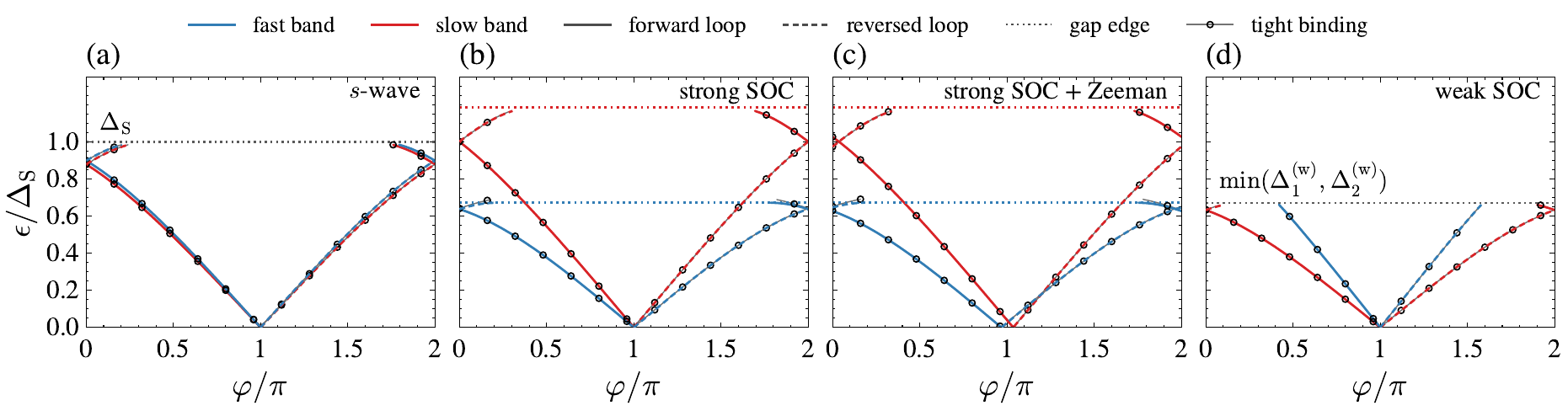}
    \caption{Andreev bound state energies as a function of the phase difference~$\varphi$, for a junction with length $L \sim \xi$. Lines show the \mbox{(quasi-)analytical}
    bound-state conditions obtained from solving Eqs.~(\ref{eq:ABS_strong}), (\ref{eq:ABS_strong_B}) or (\ref{eq:ABS_transcendental}), and open circles show the exact spectrum from a real-space tight-binding discretization of the BdG
    Hamiltonian~\eqref{eq:BdG} (see App.~\ref{app:tb}). Solid and dashed
    lines belong to the two orientations of the Andreev loop, whose levels are Kramers partners. Dotted lines mark the
    gap edges.
    The plot parameters are given in Table \ref{tab:tb} in Appendix~\ref{app:tb}.
    (a--c)~Strong spin-orbit coupling, $E_{\rm so}/\Delta_{\rm S} = 14$.
    (a)~Shows the case of pure $s$-wave pairing~($\Delta_{\mathrm{P}} = 0$) and (c) additionally has a Zeeman field $B_\parallel / \Delta_{\rm S} = 0.1$ along $\hat{\boldsymbol{m}}_{\mathrm{so}}$ applied to the normal region.
    (d)~Weak spin-orbit coupling, $E_{\rm so}/\Delta_{\rm S} = 0.15$, where the colors mark the two spin branches of
    Eq.~\eqref{eq:ABS_transcendental}.}
    \label{fig:abs_energies}
\end{figure*}

We calculate the bound-state energies using a scattering matrix approach in which we decompose the
problem into two processes. The first is the free propagation of electrons and holes through the normal region,
during which they accumulate spin-dependent dynamical phases due to the spin-orbit and Zeeman fields. The second is
the Andreev reflection at the two NS interfaces, where electrons are retroreflected as holes and vice versa. The
Andreev bound-state energies then follow from a Bohr-Sommerfeld quantization argument, where one requires
constructive interference of a quasiparticle after a complete round trip through the junction, phrased as the
determinant condition~\cite{beenakkerUniversalLimitCriticalcurrent1991}
\begin{equation}\label{eq:ABS_det}
    \det\left[\hat{t}_{\mathrm{e}} \hat{r}_{\mathrm{eh}}\hat{t}_{\mathrm{h}}\hat{r}_{\mathrm{he}} - \mathbbm{1}_2\right] = 0.
\end{equation}
Here, the transfer matrices~$\hat{t}_{e,h}$ describe the propagation of a right-moving electron and its Andreev
reflected left-moving hole. Without Zeeman field they are diagonal in the band basis and simply carry the
plane-wave phases of the two bands,
\begin{equation}\label{eq:propagation}
    \hat{t}_{s} = \mathrm{diag}\big(e^{ik_{1}^{s} L},\, e^{ik_{2}^{s} L}\big), \quad
    k_{j}^{s} = sk_{\mathrm{F}j} + \frac{\epsilon}{\hbar v_{\mathrm{F}j}},
\end{equation}
with~$s = +1(-1)$ for electrons (holes). A Zeeman field adds the term~$\mathbf B_{\mathrm Z}\cdot\boldsymbol\sigma$
to Eq.~\eqref{eq:H0_lin} in the normal region. Its component~$B_\parallel=\mathbf B_{\mathrm Z}\cdot\hat{\boldsymbol m}_{\mathrm{so}}$
keeps the transfer matrices diagonal and amounts to~$\epsilon\to\epsilon-\sigma_jB_\parallel$ in
Eq.~\eqref{eq:propagation}, where~$\sigma_{1,2}=\pm1$ is the spin polarization of the band. A transverse component
mixes the two bands, as discussed below.
The scattering matrices~$\hat{r}_{he}$ and~$\hat{r}_{eh}$ describe the Andreev reflection at the interfaces~$x = L$
and~$x = 0$, respectively. Their explicit form depends on the strength of the SOC and several cases are investigated in the following
sections. We neglect normal reflection that may be caused by a Fermi velocity mismatch between the normal and
proximitized regions.

Even within the Andreev approximation, the problem does not admit a general closed-form solution, and we therefore
consider two analytically tractable limits. In the strong spin-orbit regime, interband Andreev reflection is
suppressed and the two bands decouple to leading order (Sec.~\ref{sec:strong}). In the weak spin-orbit regime,
interband processes remain relevant and the spin-orbit coupling is treated perturbatively (Sec.~\ref{sec:weak} and
App.~\ref{app:weak}). In both cases we first analyze the case~$\mathbf{B}_{\mathrm{Z}}=0$
before discussing the effects of a finite Zeeman field.

\subsection{Strong spin-orbit coupling}\label{sec:strong}

We first consider the strong-SOC regime, in which the two bands are sufficiently separated in momentum that the
crossings between an electron branch of one band and a hole branch of the other lie far outside the superconducting
gap, see~Fig.~\ref{fig:setup}(b). These crossings lie at a distance of order~$E_{\mathrm{so}}$ from the gap edge.
In the band basis, interband mixing is generated by the off-diagonal pairing
amplitude~$\abs{\Delta_{12}}\simeq\abs{\Delta_{\mathrm{P}}\sin(\theta)}$, where $\theta$ is the angle between 
the SOC axis~$\hat{\boldsymbol{m}}_{\mathrm{so}}$ and the triplet axis~$\hat{\boldsymbol{d}}$.
This mixing may then be neglected when~$\abs{\Delta_{12}}\ll E_{\mathrm{so}}$.
We note that a momentum-independent~$\theta$ is an approximation, since in a microscopic system both axes generally depend on~$k$ and thus differ between the Fermi points of the two bands, see Table~\ref{tab:ge}.

\paragraph*{Reflection matrices and bound-state condition.}
Neglecting interband pairing, Andreev reflection is then approximately diagonal in the band basis, and each band~$j$ is gapped by the
projection of the pairing matrix onto its spin polarization~$\pm\hat{\boldsymbol{m}}_{\mathrm{so}}$,
\begin{equation}\label{eq:helicity_gaps}
    \Delta_j^{(\mathrm{s})} = \Delta_{\mathrm{S}} \pm \cos(\theta)\,\frac{k_{\mathrm{F}j}}{k_{\mathrm{F}}}\,\Delta_{\mathrm{P}},
\end{equation}
with the upper (lower) sign referring to~$j=1 (2)$.
Consistent with the Andreev approximation, we have set $k\approx\pm k_{\mathrm{F}j}$ in Eq.~(\ref{eq:helicity_gaps}) for the two bands, with
$k_{\mathrm{F}j}>0$ and the sign distinguishing right- and left-movers. Right-movers of band~$j$ are polarized
along~$\pm\hat{\boldsymbol{m}}_{\mathrm{so}}$ and left-movers, their Kramers partners, along~$\mp\hat{\boldsymbol{m}}_{\mathrm{so}}$.
Since the triplet term is odd in both momentum and spin, the two sign changes cancel, and
Eq.~\eqref{eq:helicity_gaps} holds for right- and left-movers alike.
The Andreev reflection amplitudes of the two decoupled channels thus become simply
\begin{align}\label{eq:r_strong}
    \hat{r}_{\mathrm{he/eh}} &= \diag(e^{-i\gamma_{1}}, e^{-i\gamma_{2}})\,e^{\mp i\varphi_{\mathrm{R/L}}},
\end{align}
with~$\gamma_{j} = \arccos ( \epsilon/\Delta_{j}^{(\mathrm{s})})$, where the upper (lower) sign corresponds to~$\hat{r}_{\mathrm{he}}$ ($\hat{r}_{\mathrm{eh}}$). With this, the determinant condition~\eqref{eq:ABS_det} then factorizes
into the usual transcendental equations,
\begin{equation}\label{eq:ABS_strong}
    2\pi n = \varphi + \frac{2\epsilon L}{\hbar v_{\mathrm{F}j}}
    - 2\gamma_j,
    \quad j = 1,2,
\end{equation}
where $\varphi\equiv\varphi_{\mathrm{L}}-\varphi_{\mathrm{R}}$.
Each band is thus gapped by a single helicity-locked gap during the entire round trip, in close analogy to the
Andreev states of junctions between noncentrosymmetric superconductors~\cite{asanoJosephsonEffectNoncentrosymmetric2011}.

Equation~\eqref{eq:ABS_strong} describes one orientation of the Andreev loop. The loop of reversed orientation
follows from~$\varphi\to-\varphi$. Its levels are the Kramers partners of the levels above, as required by
time-reversal symmetry. The two orientations are therefore degenerate at~$\varphi=0$ and~$\varphi=\pi$.

Figure~\ref{fig:abs_energies}(b) compares the bound-state energy as calculated from Eq.~\eqref{eq:ABS_strong} (solid curves) with a tight-binding calculation (points, see App.~\ref{app:tb} for details).
The strength of the SOC was set to $E_{\rm so}/\Delta_{\rm S} = 14$ and we used $\Delta_{\rm P}/\Delta_{\rm S} = 0.33$. All other parameters are specified in App.~\ref{app:tb}.
Figure~\ref{fig:abs_energies}(a) shows the pure
$s$-wave case, for reference. The analytic and numerical results agree well.

\paragraph*{Low-energy spectrum and level splitting.}
Taking the cosine of Eq.~\eqref{eq:ABS_strong} and expanding to linear order in the dynamical 
phase~$\epsilon L/\hbar v_{\mathrm{F}j}$ yields the closed-form approximation~\cite{parkAndreevSpinQubits2017}
\begin{equation}\label{eq:approx_strong}
    \epsilon_j \simeq \frac{\Delta_j^{(\mathrm{s})}\cos\left(\varphi/2\right)}
    {1 + \kappa_j\sin\left(\varphi/2\right)},
    \qquad
    \kappa_j = \frac{L\Delta_j^{(\mathrm{s})}}{\hbar v_{\mathrm{F}j}},
\end{equation}
where~$\kappa_j$ captures the finite-length correction characterized by the dwell time~$\tau_j =L/v_{\mathrm{F}j}$.
This approximation is accurate for short junctions,~$\kappa_j\ll1$, at all phases, and for arbitrary junction lengths
near~$\varphi=\pi$, where the bound-state energies are small.

We define the level splitting~$\hbar\omega_q=\abs{\epsilon_1-\epsilon_2}$ as the difference between the lowest positive-energy levels of the two helicity channels for a fixed orientation of the Andreev loop. Taking the difference
of the two solutions of Eq.~\eqref{eq:approx_strong} gives
\begin{equation}\label{eq:splitting_strong}
    \hbar\omega_q\simeq\frac{1}{K}\Big|
    \underbrace{\Delta^{(\mathrm s)}_{-}\cos\frac{\varphi}{2}}_{\text{gap difference}}
    -\underbrace{\frac{\Delta_1^{(\mathrm s)}\Delta_2^{(\mathrm s)}(\tau_1 - \tau_2)}{2\hbar}\sin\varphi}_{\text{velocity mismatch}}\Big|,
\end{equation}
with~$\Delta^{(\mathrm s)}_{-}=\Delta_1^{(\mathrm s)}-\Delta_2^{(\mathrm s)}
=2\cos\theta\,(\bar k_{\mathrm F}/k_{\mathrm F})\Delta_{\mathrm P}$ and~$K=[1+\kappa_1\sin(\varphi/2)][1+\kappa_2\sin(\varphi/2)]$.
The first term stems from the helicity-locked gap difference. It is first order in the triplet amplitude~$\Delta_{\mathrm{P}}$
and vanishes for~$\hat{\boldsymbol{d}}\perp\hat{\boldsymbol{m}}_{\mathrm{so}}$, where the two
helicity channels see equal gaps. The second term stems from the Fermi-velocity mismatch of the two bands and vanishes for equal velocities. The finite junction length enters through~$K$, which reduces both terms. For~$L\to0$
the velocity term and the length dependence of~$K$ vanish, while the gap difference remains. This contrasts with pure~$s$-wave pairing, where the splitting requires a finite junction length. For longer junctions
and phases away from~$\varphi=\pi$, where the levels approach the gap, the splitting follows from the full condition~\eqref{eq:ABS_strong}.

\paragraph*{Gap and velocity assignment.}
In our effective model, the triplet amplitude and the spin-orbit coupling are independent parameters.
Which band carries the larger gap is therefore not tied to which band is faster.
We therefore compare both assignments in Fig.~\ref{fig:splitting}(a). Their difference is most transparent
near~$\varphi=\pi$, where Eq.~\eqref{eq:approx_strong} becomes
\begin{equation}\label{eq:slope_times}
    \epsilon_j\simeq\frac{\hbar(\pi-\varphi)}{2\left(\tau_j+\tau_{\mathrm{A}j}\right)},
    \qquad \tau_{\mathrm{A}j}=\frac{\hbar}{\Delta_j^{(\mathrm{s})}}.
\end{equation}
The slope of each level is thus set by its total round-trip time. It combines the dwell time~$\tau_j$ in the
normal region with the Andreev reflection time~$\tau_{\mathrm{A}j}$.
If the faster band also carries the larger gap, both times are shorter in this band than in the slower one.
The two contributions then add up and enhance the splitting. In the opposite assignment, the shorter dwell time of
the faster band competes with its longer Andreev reflection time. The splitting is then reduced.
For a specific material, the assignment follows from its band structure, see Sec.~\ref{sec:ge} for a specific assessment of germanium in this context. 

Figure~\ref{fig:splitting}(a) compares the phase dependence of the level splitting for both gap assignments with the tight-binding calculation.
The solid curves follow from Eq.~\eqref{eq:ABS_strong} and the dashed curves show the approximate splitting of Eq.~\eqref{eq:splitting_strong}.
Figure~\ref{fig:splitting_soc}(a) shows the SOC dependence of the level splitting at~$\varphi=0.8\pi$, where the solid curve again follows from Eq.~\eqref{eq:ABS_strong}. The oscillations of the tight-binding points
around the analytical curve in Fig.~\ref{fig:splitting_soc}(a) stem from the interband
pairing~$\Delta_{\mathrm P}\sin\theta$, which Eq.~\eqref{eq:ABS_strong} neglects. They follow the spin-precession
angle accumulated in the normal region and disappear for~$\theta=0$.

\paragraph*{Zeeman field.}
Adding a Zeeman field leaves the helicity structure intact as long as~$\abs{{\bf B}_{\mathrm{Z}}}\ll E_{\mathrm{so}}$. 
In this regime, only the
projection~$B_\parallel = \mathbf{B}_{\mathrm{Z}}\cdot\hat{\boldsymbol{m}}_{\mathrm{so}}$ matters, given that the effect of the transverse
component will be suppressed by~$B_\perp/E_{\mathrm{so}}$. It shifts the electron and hole Fermi points of each helicity
band in opposite directions, which amounts to the replacement~$\epsilon\rightarrow\epsilon - \sigma_j B_\parallel$ in
the dynamical phase,
\begin{equation}\label{eq:ABS_strong_B}
    2\pi n = \varphi + \frac{2(\epsilon - \sigma_j B_\parallel) L}{\hbar v_{\mathrm{F}j}} - 2\gamma_j,
\end{equation}
with~$\sigma_{1,2} = \pm 1$ the spin polarization of the band. The Zeeman field thus enters only as a
band-dependent phase shift. Equation~\eqref{eq:ABS_strong_B} equals Eq.~\eqref{eq:ABS_strong} with
$\varphi\to\varphi-\sigma_j\varphi_{0,j}$, where~$\varphi_{0,j}=2B_\parallel L/\hbar v_{\mathrm{F}j}$, such that the
results above carry over to each helicity family. The two families shift in opposite directions in~$\varphi$,
lifting the remaining degeneracies of the spectrum, see~Fig.~\ref{fig:abs_energies}(c).

\subsection{Weak spin-orbit coupling}\label{sec:weak}

\begin{figure}[!t]
    \centering
    \includegraphics[width=\columnwidth]{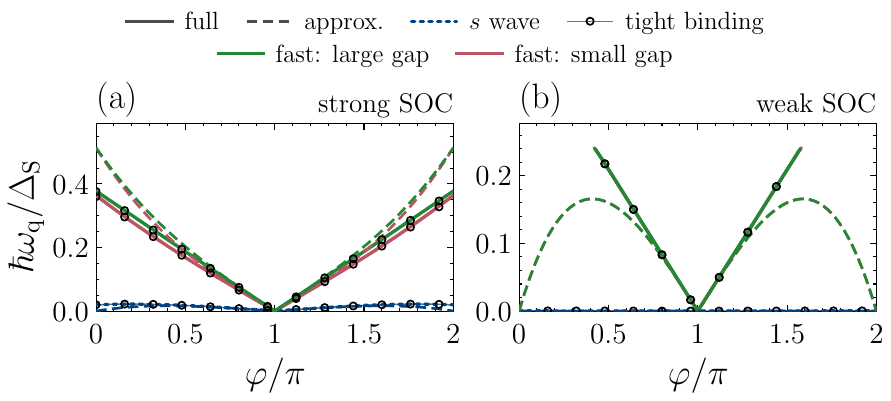}
    \caption{Phase dependence of the level splitting~$\hbar\omega_{\mathrm{q}}$ for (a)~strong and (b)~weak spin-orbit
    coupling. The parameters are those of Fig.~\ref{fig:abs_energies}. The legend refers to the gap of the faster band.
    Green lines place the larger gap on the faster band, red lines on the slower band. The two cases differ by the sign
    of the triplet component and coincide in~(b), where the velocity mismatch is negligible. Blue dotted lines show the
    pure $s$-wave reference with equal gaps, split only by the Fermi-velocity mismatch. Solid curves follow from
    Eq.~\eqref{eq:ABS_strong} in~(a) and from Eq.~\eqref{eq:ABS_transcendental} with reflection matrices up to second
    order in~(b). Dashed curves show the approximate splittings of Eq.~\eqref{eq:splitting_strong} in~(a) and
    Eq.~\eqref{eq:splitting_weak} in~(b). Black lines with open circles show the corresponding tight-binding results.
    In~(a), each level is followed up to the gap of its own band. In~(b), the curves end where the upper level reaches
    the smaller of the two gaps.
    In both, the finite splitting at~$\varphi=0$ is consistent with time-reversal symmetry, since $\hbar\omega_q$ separates levels of different helicity bands rather than Kramers partners.}
    \label{fig:splitting}
\end{figure}

We now consider the weak-SOC regime, where we treat the spin-orbit coupling and the Fermi velocity mismatch perturbatively.
We assume that the crossings between an electron branch of one band and a hole branch of the other lie within the superconducting
pairing window, see~Fig.~\ref{fig:setup}(c). Interband pairing can therefore no longer be neglected, and the full
spin structure of the pairing matrix must be retained. We evaluate the pairing at the common Fermi
momentum~$k_{\mathrm{F}}$ of the underlying parabolic band and define
$\Delta_{1,2}^{(\mathrm{w})}=\Delta_{\mathrm{S}}\pm\Delta_{\mathrm{P}}$.

A perturbative approach requires $\abs{E_{\mathrm{so}}}\ll\Omega_j$ for both bands, where $\Omega_j=\sqrt{(\Delta_j^{(\mathrm{w})})^2-\epsilon^2}$ sets the decay of the quasiparticles
into the superconductor, see App.~\ref{app:weak_perturbative}. The expansion thus holds for
$E_{\mathrm{so}}\ll\Delta_{1,2}^{(\mathrm{w})}$ and fails only close to the gap edges.
An essential feature of weak SOC is that Andreev reflection generally mixes the two spin-orbit bands. An incident
particle is reflected into a superposition of the two outgoing channels. This spin mixing is produced by
triplet pairing misaligned with the SOC axis.

\paragraph*{Reflection matrices and bound-state condition.} Below the gap the Andreev reflection conserves the
current, so the reflection matrices are unitary. In the band basis,
where~$\hat{\boldsymbol m}_{\mathrm{so}}\cdot\boldsymbol\sigma=\sigma_z$, they can therefore be written as
\begin{equation}\label{eq:r_weak}
    \hat r_{\mathrm{he/eh}}=
    e^{-i(\chi+\lambda\hat{\boldsymbol b}_\pm\cdot\boldsymbol\sigma)}
    e^{\mp i\varphi_{\mathrm{R}/\mathrm{L}}},
\end{equation}
with the upper (lower) sign referring to~$\hat r_{\mathrm{he}}$ ($\hat r_{\mathrm{eh}}$), as in
Eq.~\eqref{eq:r_strong}. The common phase~$\chi$ describes the spin-independent Andreev phase shift. The
vector~$\hat{\boldsymbol b}_\pm$ is the axis of a spin rotation by the angle~$2\lambda$.

Between the interfaces the transfer matrices~\eqref{eq:propagation} separate into a common phase and a spin rotation about the SOC axis,
\begin{equation}\label{eq:weak_transfer}
    \hat t_{e,h}=e^{\pm i\bar k_{\mathrm F}L+i\epsilon L/(\hbar\bar v)}\,e^{i(\delta_\epsilon\mp\Phi)\sigma_z}.
\end{equation}
Here~$\bar v=2v_{\mathrm{F}1}v_{\mathrm{F}2}/(v_{\mathrm{F}1}+v_{\mathrm{F}2})$ is the harmonic-mean Fermi velocity.
The angle~$\Phi=(k_{\mathrm F2}-k_{\mathrm F1})L/2=E_{\mathrm{so}}L/(\hbar\bar v)$ is the spin precession caused by
the SOC in the normal region. The energy-dependent spin phase~$\delta_\epsilon=\epsilon(\tau_1-\tau_2)/2\hbar$ stems
from the velocity mismatch. Inserting Eqs.~\eqref{eq:r_weak} and~\eqref{eq:weak_transfer} into the determinant
condition~\eqref{eq:ABS_det} yields (see also App.~\ref{app:weak_abs})
\begin{equation}\label{eq:ABS_transcendental}
    2\pi n=\varphi+\frac{2\epsilon L}{\hbar\bar v}-2\chi+\eta\zeta,
    \qquad \eta=\pm1,
\end{equation}
The angle~$\zeta$ is the net spin rotation over a complete Andreev round trip. It follows from
\begin{equation}\label{eq:zeta_trace}
    \cos\zeta=\tfrac12\operatorname{Tr}\mathcal W,
\end{equation}
where~$\mathcal W=e^{i(\delta_\epsilon-\Phi)\sigma_z}e^{-i\lambda\hat{\boldsymbol b}_-\cdot\boldsymbol\sigma}
e^{i(\delta_\epsilon+\Phi)\sigma_z}e^{-i\lambda\hat{\boldsymbol b}_+\cdot\boldsymbol\sigma}$ is the spin-dependent part of the
round trip~$\hat t_e\hat r_{\mathrm{eh}}\hat t_h\hat r_{\mathrm{he}}$. The two eigenchannels of~$\mathcal W$ acquire
opposite phases~$\pm\zeta$, which gives the two spin branches~$\eta=\pm1$.

\paragraph*{Spin-orbit-free leads.} A consistent perturbative treatment that includes the leading-order effect of spin-orbit coupling in both the superconducting and normal regions requires~$\chi$, $\lambda$
and~$\hat{\boldsymbol b}_\pm$ to second order in~$E_{\mathrm{so}}$ and~$\delta v_{\mathrm{F}}/v_0$, which we derive
in App.~\ref{app:weak_parameters}.
However, it is instructive to first consider the leading-order result in which SOC and the velocity mismatch inside the superconductor are neglected.
This approximation is useful to describe systems with spin-orbit-free superconducting leads or with~$L \gg \xi$ (see below), such that the spin-orbital mixing in the leads can be neglected compared to that in the normal region.
In this case, the spin components parallel and antiparallel to~$\hat{\boldsymbol d}$ then see the gaps~$\Delta_{1,2}^{(\mathrm{w})}$ and acquire the Andreev phases~$\alpha_j=\arccos(\epsilon/\Delta_j^{(\mathrm{w})})$.
In the band basis this gives
\begin{equation}\label{eq:r_weak0}
    \hat r_{\mathrm{he/eh}}^{(0)} =
    e^{-i(\alpha_++\alpha_-\hat{\boldsymbol d}\cdot\boldsymbol\sigma)}
    e^{\mp i\varphi_{\mathrm{R}/\mathrm{L}}},
\end{equation}
with~$\alpha_\pm=(\alpha_1\pm\alpha_2)/2$. This corresponds to~$\chi=\alpha_+$, $\lambda=\alpha_-$
and~$\hat{\boldsymbol b}_\pm=\hat{\boldsymbol d}$ in Eq.~\eqref{eq:r_weak}. The reflection rotates the spin about the
triplet axis by the angle~$2\alpha_-$. If~$\hat{\boldsymbol d}$ is not aligned with~$\hat{\boldsymbol m}_{\mathrm{so}}$,
this rotation mixes the two bands. Equation~\eqref{eq:zeta_trace} then takes the simple form
\begin{align}\label{eq:zeta0}
    \begin{split}
        \cos\zeta=2&\left(\cos\alpha_-\cos\delta_\epsilon+\sin\alpha_-\cos\theta\sin\delta_\epsilon\right)^2 \\
        &+2\sin^2\alpha_-\sin^2\theta\sin^2\Phi-1.
    \end{split}
\end{align}
The gap asymmetry enters through~$\alpha_-$, the velocity mismatch through~$\delta_\epsilon$ and the SOC in the
normal region through~$\Phi$. For equal gaps Eq.~\eqref{eq:zeta0} reduces to~$\cos\zeta=\cos2\delta_\epsilon$, the
known result for $s$-wave pairing with
SOC~\cite{chtchelkatchevAndreevQuantumDots2003,parkAndreevSpinQubits2017,tosiSpinOrbitSplittingAndreev2019}.
For leads without SOC, Eq.~\eqref{eq:zeta0} is exact. For leads with SOC, it neglects the spin precession inside the
superconductor, which acts over the coherence length~$\xi$. For~$\Phi\ll1$ this precession compares to that in the normal region as~$\xi/L$, so that Eq.~\eqref{eq:zeta0} applies to junctions with~$L\gg\xi$. For longer junctions~$\Phi$ enters only periodically. The lead contribution then becomes relevant again near the values of~$\Phi$ where the normal-region contribution vanishes.

\paragraph*{Low-energy spectrum with spin-orbit-free leads.}
For small positive energies,~$\zeta\simeq\epsilon\zeta_1$ in Eq.~\eqref{eq:ABS_transcendental}, where
\begin{equation}\label{eq:zeta1}
    \zeta_1=\sqrt{(g_v-g_\parallel)^2+g_\perp^2}
\end{equation}
describes the linear energy dependence of the round-trip spin phase.
Here~$g_v=(\tau_1-\tau_2)/\hbar$ is the contribution of the Fermi-velocity mismatch and the coefficients $g_\parallel$ and $g_\perp$ describe the pairing contribution to the energy-dependent spin rotation along and perpendicular to the SOC axis.
They account for both Andreev reflections and the spin precession between the interfaces.
The longitudinal contribution can reinforce or cancel the velocity contribution.
The transverse contribution acts along an orthogonal spin axis and cannot cancel it.

When neglecting the SOC in the leads, i.e., using Eq.~(\ref{eq:zeta0}), we can approximate $\chi\simeq\arccos(\epsilon/\Delta_{\mathrm{eff}})$, with the
effective gap $\Delta_{\mathrm{eff}}=(\Delta_{\mathrm S}^2-\Delta_{\mathrm P}^2)/\Delta_{\mathrm S}$, and
\begin{align}
    g_\parallel = {} & {} \frac{2\Delta_{\rm P}}{\Delta_{\rm S}^2 - \Delta_{\rm P}^2} \cos\theta,\\
    g_\perp = {} & {} \frac{2\Delta_{\rm P}}{\Delta_{\rm S}^2 - \Delta_{\rm P}^2}\sin\theta\cos\Phi.
\end{align}
We see that the transverse contribution to the spin rotation in this case oscillates with~$E_{\mathrm{so}}L$ and vanishes for~$\Phi=(2n+1)\frac{\pi}{2}$.

We now insert the low-energy expressions for~$\chi$ and~$\zeta$ into Eq.~\eqref{eq:ABS_transcendental}. Expanding to
first order in the energy-dependent propagation and spin phase shifts gives the approximate spectrum
\begin{equation}\label{eq:ABS_approx}
 \epsilon_\eta\simeq
 \frac{\Delta_{\mathrm{eff}}\cos(\varphi/2)}
 {1+\kappa_\eta\sin(\varphi/2)},\qquad
 \kappa_\eta=\bar\kappa+\eta\frac{\Delta_{\mathrm{eff}}\zeta_1}{2},
\end{equation}
with~$\bar\kappa=L\Delta_{\mathrm{eff}}/(\hbar\bar v)$. This has the same form as the strong-SOC
approximation~\eqref{eq:approx_strong}, but the two branches now depend on the combined round-trip spin rotation. It
describes the low-energy spectrum near~$\varphi=\pi$. 
Taking the difference of the two solutions gives
\begin{equation}\label{eq:splitting_weak}
    \hbar\omega_q\simeq
    \frac{\Delta_{\mathrm{eff}}^2}{2K}\,\zeta_1\abs{\sin\varphi},
\end{equation}
where~$K$ is defined as below Eq.~\eqref{eq:splitting_strong}, with~$\kappa_{1,2}$ replaced by~$\kappa_\pm$.
The cross term~$-2g_vg_\parallel$ in Eq.~\eqref{eq:zeta1} determines whether the longitudinal pairing and velocity
contributions reduce or enhance the splitting. The full transcendental condition~\eqref{eq:ABS_transcendental}
needs to be considered as the levels approach the continuum edge.

\paragraph*{Fully spin-orbit-coupled junction.} If the superconducting regions also have SOC, the spin precesses while the
quasiparticle penetrates the superconductor over its decay length~$\hbar v_0/\Omega_j$. In App.~\ref{app:weak_parameters} we derive the leading-order contributions to $\lambda$, $\chi$, and ~$\hat{\boldsymbol b}_\pm$ that capture the spin precession
inside the superconductor. 
The SOC component along~$\hat{\boldsymbol d}$ shifts electron and hole momenta equally and leaves
the Andreev phases unchanged, while the perpendicular component mixes the two pairing channels. Like an off-diagonal
perturbation, it tilts~$\hat{\boldsymbol b}_\pm$ out of the plane spanned by~$\hat{\boldsymbol d}$
and~$\hat{\boldsymbol m}_{\mathrm{so}}$ already at first order, by an angle of order~$E_{\mathrm{so}}/\Omega_j$ with
opposite signs at the two interfaces. In~$\cos\zeta$ this tilt appears together with the precession~$\Phi$ in the
normal region, see App.~\ref{app:weak_abs}. 
For~$\Phi\ll1$ the two compare as~$\xi/L$, which is why Eq.~\eqref{eq:zeta0}
is also valid at~$L\gg\xi$. The tilt also keeps~$g_\perp$ finite at~$\Phi=\pi/2$, with a residual of relative
order~$E_{\mathrm{so}}/\Delta_{\mathrm S}$. The Andreev phases~$\chi\pm\lambda$, in contrast, shift only at second
order in $E_{\rm so}$, so that second order is the lowest order at which all parameters are treated consistently. The corrections
to~$\chi$ modify the effective gap,
\begin{equation}\label{eq:Delta_eff_weak}
 \Delta_{\mathrm{eff}}\simeq
 \frac{\Delta_{\mathrm{S}}^2-\Delta_{\mathrm{P}}^2}{\Delta_{\mathrm{S}}}
 +\frac{E_{\mathrm{so}}^2\Delta_{\mathrm{P}}^2}{\Delta_{\mathrm{S}}^3}\sin^2\theta,
\end{equation}
whereas the velocity mismatch does not change~$\Delta_{\mathrm{eff}}$ at this order. The low-energy
results~\eqref{eq:ABS_approx} and~\eqref{eq:splitting_weak} keep their form, but one needs to use the $\Delta_{\mathrm{eff}}$ given in Eq.~(\ref{eq:Delta_eff_weak}) and coefficients~$g_\parallel$ and~$g_\perp$ in~$\zeta_1$ that are also corrected to second order in $E_{\rm so}$, see App.~\ref{app:weak_low_energy}.

\begin{figure}[t!]
    \centering
    \includegraphics[width=\columnwidth]{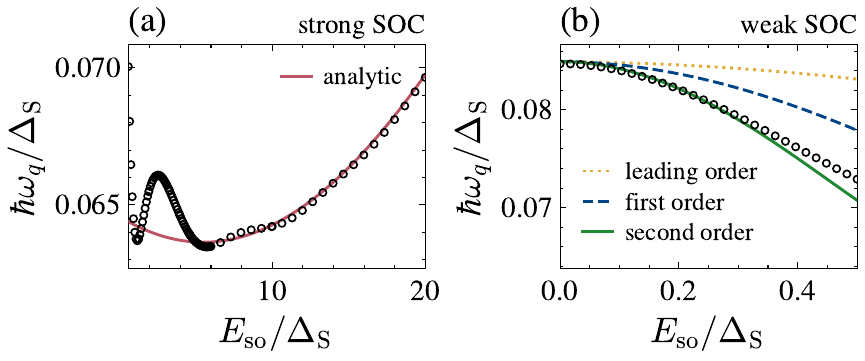}
    \caption{SOC dependence of the splitting between the two lowest positive-energy ABS levels at
    $\varphi=0.8\pi$. All other parameters are those of Fig.~\ref{fig:splitting}. 
    Open circles show the tight-binding results. In~(a), the red solid line follows from the strong-SOC
    condition~\eqref{eq:ABS_strong}. In~(b), the yellow dotted, blue dashed and green solid lines follow from the
    weak-SOC condition~\eqref{eq:ABS_transcendental} using reflection matrices at leading, first and second order
    in~$E_{\mathrm{so}}$ and~$\delta v_{\mathrm F}/v_0$, respectively.}
    \label{fig:splitting_soc}
\end{figure}

In Fig.~\ref{fig:abs_energies}(d) we show an example of the bound-state spectrum for the case of weak spin-orbit coupling [with $E_{\rm so}/\Delta_{\rm S} = 0.15$ and otherwise the same parameters as in Fig.~\ref{fig:abs_energies}(b)], found from solving Eq.~(\ref{eq:ABS_transcendental}) using the second-order expressions for $\chi$ and $\zeta$ given in App.~\ref{app:weak_parameters}.
The open circles present again the results from a tight-binding calculation, detailed in App.~\ref{app:tb}.
Figure~\ref{fig:splitting}(b) compares the
phase-dependent splitting for the two gap assignments with the pure $s$-wave reference, where the solid curves follow from Eq.~\eqref{eq:ABS_transcendental} with reflection matrices up to second order in~$E_{\rm so}$ and the dashed curves show the approximate splittings of Eq.~\eqref{eq:splitting_weak}. For weak SOC, the two bands
have nearly equal velocities and~$g_v\approx0$. The splitting then stems from the triplet pairing alone. Reversing
the sign of~$\Delta_{\mathrm P}$ leaves it unchanged, since the cross term in Eq.~\eqref{eq:zeta1} vanishes.
Figure~\ref{fig:splitting_soc}(b) shows the SOC dependence of the level splitting. The first- and second-order
corrections progressively improve the agreement with the tight-binding calculation as~$E_{\mathrm{so}}$ grows.

\paragraph*{Effect of a Zeeman field.} Since we assume the Zeeman field to only act in the normal region, it leaves the second-order reflection parameters
$\chi$, $\lambda$, and~$\hat{\boldsymbol b}_\pm$ unchanged. It modifies the spin precession between the interfaces.
For equal band velocities, the electron and hole precession is governed by the combined fields
\begin{equation}\label{eq:Bfields}
    \mathbf{\mathcal{B}}_{e,h} = \mathbf{B}_{\mathrm{Z}} \pm E_{\mathrm{so}}\hat{\boldsymbol{m}}_{\mathrm{so}},
\end{equation}
whose axes need not be collinear. Combining these propagation rotations with the corrected interface rotations gives
a field-dependent round-trip angle~$\zeta_B$. For unequal velocities, the field also produces a common propagation
phase~$\delta\varphi_B=2aB_\parallel L/(\hbar\bar v)$, where~$a=\delta v_{\mathrm{F}}/v_0$ and
$B_\parallel=\mathbf B_{\mathrm Z}\cdot\hat{\boldsymbol m}_{\mathrm{so}}$.
The bound-state condition becomes
\begin{equation}\label{eq:ABS_weak_second_order_B_main}
 2\pi n=\varphi+\frac{2\epsilon L}{\hbar\bar v}
 +\delta\varphi_B-2\chi+\eta\zeta_B,
\end{equation}
which can again be solved numerically, or explicitly for small energies.
The general propagation matrices and~$\zeta_B$ are derived in App.~\ref{app:weak_zeeman}.

\section{Proximitized germanium nanowire}\label{sec:ge}

The preceding analysis treats the Fermi momenta, the Fermi velocities and the pairing matrix as effective model
parameters. To connect to a concrete platform, we consider in this section a germanium hole wire and use numerical simulations to obtain more realistic estimates for the parameters. In this system, the combination of
confinement, spin-orbit coupling and the mixing of heavy and light holes generates band-dependent velocities. We
obtain~$k_{\mathrm{F}j}$ and~$v_{\mathrm{F}j}$ by diagonalizing a three-dimensional tight-binding discretization of
the Luttinger-Kohn Hamiltonian~\cite{luttingerQuantumTheoryCyclotron1956, winklerSpinOrbitCoupling2003,
willatzenMethodElectronicProperties2009}. In the spherical approximation~$\gamma_2=\gamma_3\equiv\gamma_s$, it
reads
\begin{equation}
    H_{\rm LK}=\frac{\hbar^2}{2m_h}\Big[\Big(\gamma_1+\tfrac52\gamma_s\Big)k^2-2\gamma_s(\bm k\cdot\bm J)^2\Big],
\end{equation}
where~$m_h=-m_0$ is the bare hole mass and~$J_{x,y,z}$ are the spin-$3/2$ matrices. We work in the angular momentum basis
$\{\ket{\frac{3}{2},\frac{3}{2}},\ket{\frac{3}{2},-\frac{3}{2}},\ket{\frac{3}{2},\frac{1}{2}},\ket{\frac{3}{2},-\frac{1}{2}}\}$ with quantization axis~$z$. The states
with~$m_J=\pm\frac{3}{2}$ ($\pm\frac{1}{2}$) are the heavy (light) holes.

We consider a quasi-one-dimensional rectangular wire and take parameters for Ge, see App.~\ref{app:ge} for details. Spin-orbit coupling is taken into account via direct Rashba coupling~\cite{kloeffelStrongSpinorbitInteraction2011, kloeffelDirectRashbaSpinorbit2018, winklerSpinOrbitCoupling2003}, by introducing a homogeneous electric field along the $y$-direction, perpendicular to the length direction of the wire. 
To find the effective pairing, we use a pairing matrix $\Delta_{\rm LK}$ (using the notation from Ref.~\cite{babkinSuperconductingProximityEffect2025}) in the time-reversed/BdG basis and project the proximity-induced pairing onto the two highest
valence subbands. The resulting effective~$2\times2$ pairing matrix then provides the input for the one-dimensional model studied above.

\begin{table}[t]
\begin{ruledtabular}
\begin{tabular}{lcccc}
 & \multicolumn{2}{c}{unstrained} & \multicolumn{2}{c}{strained} \\
 & $j=1$ & $j=2$ & $j=1$ & $j=2$ \\
\hline
$k_{\mathrm{F}j}$ (nm$^{-1}$) & $0.077$ & $0.144$ & $0.066$ & $0.136$ \\
$v_{\mathrm{F}j}$ ($10^{5}$~m/s) & $1.36$ & $1.62$ & $1.49$ & $1.61$ \\
$\Delta^{(\mathrm s)}_j$ ($\mu$eV) & $107.6$ & $86.8$ & $55.8$ & $66.8$ \\
\hline
$\Delta_{\mathrm S}$ ($\mu$eV) & $98$ & $99$ & $60$ & $58$ \\
$\Delta_{\mathrm P}$ ($\mu$eV) & $15$ & $29$ & $7$ & $13$ \\
$\theta$ & $48^\circ$ & $65^\circ$ & $58^\circ$ & $46^\circ$ \\
$E_{\mathrm{so}}$ (meV) & \multicolumn{2}{c}{$3.3\approx34\,\Delta_{\mathrm S}$} & \multicolumn{2}{c}{$3.6\approx61\,\Delta_{\mathrm S}$} \\
$\abs{\Delta_{12}}/E_{\mathrm{so}}$ & \multicolumn{2}{c}{$\approx8\times10^{-3}$} & \multicolumn{2}{c}{$\approx2\times10^{-3}$} \\
$\bar\xi$ ($\mu$m) & \multicolumn{2}{c}{$\approx1.0$} & \multicolumn{2}{c}{$\approx1.7$} \\
\end{tabular}
\end{ruledtabular}
\caption{Effective parameters of the simulated Ge wire without and with strain, at a Fermi energy of $6.2$~meV (unstrained) and $5.7$~meV (strained) below the top of the highest subband. Band~2 has the larger Fermi
momentum. The helicity gaps~$\Delta^{(\mathrm s)}_j$ are the diagonal pairing elements found at the Fermi point of each
band. $\Delta_{\mathrm S}$, $\Delta_{\mathrm P}$ and~$\theta$ follow from writing the projected pairing as
$\Delta_{\mathrm S}+\Delta_{\mathrm P}\hat{\boldsymbol d}\cdot\boldsymbol\sigma$ in the band basis. They are evaluated at the Fermi point of band~$j$. The last two rows give the ratio of the interband pairing to the spin-orbit energy and the coherence length~$\bar\xi=\hbar\bar v_{\mathrm F}/\Delta_{\mathrm S}$.}
\label{tab:ge}
\end{table}

Table~\ref{tab:ge} lists the resulting parameters for the wire with and without strain.
The strain is compressive in the $xy$-plane and tensile along~$z$, see App.~\ref{app:ge} for the strain model. Both parameter sets are
evaluated about $6$~meV below the top of the highest subband. In both cases the spin-orbit energy exceeds the induced
gap by a factor of 30 to 60. The interband pairing is more than two orders of magnitude smaller than the spin-orbit energy, which sets the energy
separation of the two bands at the Fermi level. This Ge wire therefore lies deep in the strong-SOC regime of Sec.~\ref{sec:strong}, where the splitting is
set by the two helicity gaps and the two Fermi velocities.

We see that the magnitude of the induced gap depends significantly on strain, which is caused by a strongly strain-dependent orbital content of the eigenstates in the semiconductor.
As in the pairing model of Ref.~\cite{babkinSuperconductingProximityEffect2025}, we characterize the coupling of the states in the superconductor to the three $p$-orbitals by the tunneling vector $\hat{\boldsymbol t}$.
For a clean interface to a two-dimensional hole gas, only $t_z$ is expected to be non-zero, yielding heavy holes that are not proximitized directly since they do not contain a $p_z$-component.
We allow for a small tilt of~$\hat{\boldsymbol t}$, as caused by an imperfect interface, but keep it predominantly along~$z$.
Strain, confinement and electric fields all affect the orbital content of the eigenstates in the wire, and we find that for the parameters chosen here adding strain significantly reduces the $p_z$-component in the lowest-energy (heavy-hole-like) eigenstates and thus yields smaller gaps than in the unstrained case.
Furthermore, the Fermi velocities of the two bands differ by $19\%$ without and $8\%$ with strain, and we find that without strain the slower band has the larger gap, by about~$20\%$, whereas the relative magnitude of the gaps is opposite in the strained case.
This means that in the unstrained case the gap asymmetry and the velocity mismatch counteract each other in Eq.~\eqref{eq:splitting_strong}, yielding a vanishing level splitting at the compensation length~$L^*\approx1.3~\mu$m.
In the strained wire, the faster band has the larger gap, and both contributions add.
Which band carries the larger gap depends on the details of the band structure, such as the subband spacing, and is not fixed by symmetry.

\begin{figure}[t]
    \centering
    \includegraphics[width=\columnwidth]{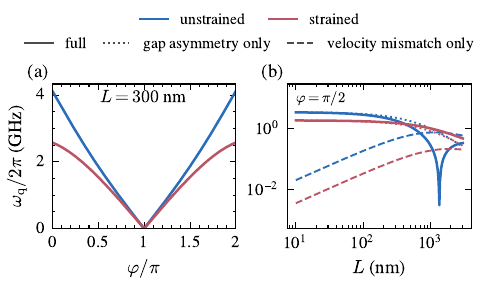}
    \caption{Level splitting of the strong-SOC Andreev doublet for the unstrained (blue) and strained (red) Ge wire
    of Table~\ref{tab:ge}, obtained from Eq.~\eqref{eq:ABS_strong}. (a)~Phase dependence for $L=300$ nm.
    (b)~Length dependence at $\varphi=\pi/2$. Dotted lines keep only the gap asymmetry (equal velocities) and dashed
    lines only the velocity mismatch (equal gaps). The full unstrained splitting vanishes at the compensation
    length~$L^*$.}
    \label{fig:ge_splitting}
\end{figure}

Inserting the helicity gaps and Fermi velocities of Table~\ref{tab:ge} into the strong-SOC bound-state condition~\eqref{eq:ABS_strong} gives the level splitting shown in Fig.~\ref{fig:ge_splitting}. At~$L=300$ nm the gap asymmetry dominates in both
wires. The splitting is then largest at~$\varphi=0$ and decreases towards~$\varphi=\pi$. Figure~\ref{fig:ge_splitting}(b) shows that the gap asymmetry favors short junctions,
while the velocity contribution grows with~$L$ up to $L\sim\bar\xi$. In the unstrained wire the two compete and the
splitting vanishes at~$L^*$. In the strained wire they add, so that its splitting exceeds the unstrained one for
$L\gtrsim400$ nm. For short junctions this agrees with Ref.~\cite{coppiniStrainEngineeringAndreev2026a}, where in-plane compressive strain also
suppresses the splitting. The mechanism differs, however. In the planar wells of that work strain reduces the mixing of
heavy and light holes and with it the spin-orbit coupling. In our narrow wire the strong electric field keeps the
spin-orbit energy above~$3$~meV, and strain reduces the splitting by lowering the gaps. In our wire the splitting therefore remains in the GHz range also with strain, whereas Ref.~\cite{coppiniStrainEngineeringAndreev2026a} finds it strongly suppressed for comparable strain. For longer junctions the
added velocity contribution can even reverse this trend.

\section{Conclusion}\label{sc:conclusion}
We have developed an analytical description of the spin-resolved Andreev spectrum in ballistic semiconductor Josephson junctions
with mixed singlet and triplet pairing. By retaining both the pairing structure and the band-dependent Fermi velocities,
we show that triplet pairing provides an additional mechanism for level splitting.
This contribution remains finite for equal velocities and can reinforce or partially compensate the splitting associated
with a velocity mismatch.

The pairing acts differently in the strong- and weak-SOC regimes. For strong SOC, interband pairing is suppressed and the
two spin-orbit bands acquire distinct projected gaps. Their difference is controlled by the orientation of the triplet pairing
relative to the SOC axis. For weak SOC, misalignment of these axes instead produces spin-mixing Andreev reflection.
We derived the reflection matrices through second order in the SOC strength and the relative velocity mismatch. We
identified how the interface rotations combine with spin precession through the normal region to determine the
bound-state spectrum.

Comparisons with real-space tight-binding calculations support the analytical descriptions within their respective regimes
of validity. In particular, the successive weak-SOC corrections improve the prediction of the SOC dependence of the level
splitting. Effective parameters extracted from a microscopic model of a proximitized germanium hole wire place it deep in
the strong-SOC regime. There the helicity gaps differ by about~$20\%$, because the induced pairing depends on the
light-hole content of each band. Without strain this gap asymmetry partly compensates the velocity mismatch. In-plane compressive strain lowers the gaps and reverses their order, so that both contributions add. Our results clarify how mixed pairing
shapes the Andreev spectrum and provide a basis for assessing its consequences for devices such as Andreev spin qubits.

\begin{acknowledgments}
This work was supported by the Research Council of Norway through Grant No.~353894, the European Research Council (Grant Agreement No.~856526) and the DFG Collaborative Research Center 183 (Project No.~277101999).
FVB acknowledges funding from the European Union's Horizon Europe research and innovation programme under the Marie Sk\l{}odowska-Curie grant agreement No.~101204715. \paragraph*{AI usage statement} 
We used OpenAI Astra~6 for the perturbation-theory calculations and Claude Opus 5.5 for assistance in reviewing the manuscript. We reviewed the resulting calculations, formatting and text. 
\end{acknowledgments}

\appendix

\section{Andreev reflection matrices for weak spin-orbit coupling}
\label{app:weak}

This appendix derives the reflection matrices of Eq.~\eqref{eq:r_weak}. We start from the linearized Hamiltonian~\eqref{eq:H0_lin} and expand in
the spin-orbit energy~$E_{\mathrm{so}}$ and in the relative velocity difference~$a=\delta v_{\mathrm F}/v_0$, which is also proportional to the spin-orbit strength. Both
are treated as small quantities of the same order, and all results are accurate through second order. The pairing is kept without expanding in~$\Delta_{\mathrm{P}}$.

\subsection{Pairing basis}\label{app:weak_setup}

For weak SOC the pairing is evaluated at the common reference momentum~$k_{\mathrm F}$. Disregarding the phase factor for now, Eq.~\eqref{eq:pairing_leads} reduces to the spin matrix~$\Delta_{\mathrm S}+\Delta_{\mathrm P}\,
\hat{\boldsymbol d}\cdot\boldsymbol\sigma$. Its eigenstates are the spin states parallel and antiparallel
to~$\hat{\boldsymbol d}$. We use them as basis and refer to it as the pairing basis. In this basis the pairing is
diagonal,
\begin{equation}\label{eq:weak_pairing_basis}
 D=\operatorname{diag}(\Delta_1,\Delta_2),\qquad
 \Delta_{1,2}\equiv\Delta_{1,2}^{(\mathrm{w})}=\Delta_{\mathrm{S}}\pm\Delta_{\mathrm{P}},
\end{equation}
with~$\Delta_{\mathrm S}>\abs{\Delta_{\mathrm P}}$, and the SOC operator~$\hat{\boldsymbol m}_{\mathrm{so}}\cdot\boldsymbol\sigma$ reads
\begin{equation}\label{eq:weak_M}
 M=\begin{pmatrix}\cos\theta& \sin\theta\\\sin\theta&-\cos\theta\end{pmatrix},
\end{equation}
where $\theta$ is the angle between $\hat{\boldsymbol d}$ and $\hat{\boldsymbol m}_{\mathrm{so}}$, which we assume to lie in the $xz$-plane.
The eigenstates of $M$ are the two spin-orbit bands,
\begin{equation}\label{eq:band_spinors}
\ket{1}=\begin{pmatrix}\cos(\theta/2)\\\sin(\theta/2)\end{pmatrix},\qquad
\ket{2}=\begin{pmatrix}-\sin(\theta/2)\\\cos(\theta/2)\end{pmatrix},
\end{equation}
with~$M\ket{1}=\ket{1}$ and~$M\ket{2}=-\ket{2}$. The unitary matrix~$W=(\ket{1},\ket{2})$ satisfies~$W^\dagger MW=\sigma_z$.
It maps band-basis amplitudes onto pairing-basis amplitudes, so that a matrix~$X$ in the pairing basis
becomes~$W^\dagger XW$ in the band basis. In the time-reversed-hole convention of Eq.~\eqref{eq:BdG} the same
rotation acts on the electron and the hole block.
The linearized BdG Hamiltonian thus reads
\begin{equation}\label{eq:HBdG_weak}
    H(x) = \big[\hbar (-i\partial_x - k_{\rm F})F+E_{\mathrm{so}}M \big]\tau_z + D\tau_x,
\end{equation}
with $F=v_0(\mathbbm{1} + aM)$, where $v_0=(v_{\mathrm F1}+v_{\mathrm F2})/2$ and~$a =(v_{\mathrm F1}-v_{\mathrm F2})/2 v_0$, see also below Eq.~\eqref{eq:H0_lin}, and $\tau_{x,z}$ are Pauli matrices in Nambu space.

\subsection{Reflection matrices}\label{app:weak_reflection}

Consider an NS interface with the superconductor to its right ($\nu=+$) or to its left ($\nu=-$). At energies below
the gap two modes decay into the superconductor. 
The superconducting wave function can thus be written as
\begin{equation}
    \Psi_{{\rm S},\nu}(x)
    = \sum_{j=1}^{2} c_{j\nu}
    \begin{pmatrix} u_{j\nu} \\ v_{j\nu} \end{pmatrix}
    e^{iq_{j\nu}x}.
\end{equation}
The two-component spinors~$u_{j\nu}$ and~$v_{j\nu}$ describe their electron and hole components, respectively.
Introducing
\begin{equation}
    U_\nu=\left( \begin{array}{cc} u_{1\ua\nu} & u_{2\ua\nu} \\
    u_{1\da\nu} & u_{2\da\nu} \end{array}\right),\quad
    V_\nu=\left( \begin{array}{cc} v_{1\ua\nu} & v_{2\ua\nu} \\
    v_{1\da\nu} & v_{2\da\nu} \end{array}\right),
\end{equation}
and $c_\nu=(c_{1\nu},c_{2\nu})^T$, the interface wave function becomes
\begin{equation}
    \Psi_{{\rm S},\nu}(0)
    =\begin{pmatrix}\psi_e\\\psi_h\end{pmatrix}
    =\begin{pmatrix}U_\nu c_\nu\\V_\nu c_\nu\end{pmatrix}.
\end{equation}
Continuity identifies these components with the incoming and outgoing
normal-region amplitudes. Focusing on the case where at the right interface an incident electron is reflected as a hole (and at the left interface an incident hole is reflected as an electron) one can eliminate $c_\nu$ straightforwardly, yielding
\begin{equation}\label{eq:weak_boundary_reflection}
    \mathcal R_+=V_+U_+^{-1}\quad(\mathrm{he}),\qquad
    \mathcal R_-=U_-V_-^{-1}\quad(\mathrm{eh}).
\end{equation}
Inserting the decaying modes into the BdG equation $[H(x)-\epsilon]\Psi_{{\rm S},\nu}(x)=0$ and eliminating their complex momenta
gives a closed equation for~$\mathcal R_\nu$,
\begin{equation}\label{eq:weak_riccati_G}
    \begin{split}
        &\mathcal R_\nu D\mathcal R_\nu - 2\epsilon\mathcal R_\nu + D
        -\nu E_{\mathrm{so}}[M,\mathcal R_\nu] \\
        &+ a \epsilon\{M,\mathcal R_\nu\} - a \mathcal R_\nu M D \mathcal R_\nu -a MD =0.
    \end{split}
\end{equation}
This quadratic matrix equation is exact within the linearized model. The physical solution is selected by requiring that the superconducting wave function decays away from the interface. We obtain it perturbatively
from the corresponding decaying solution at~$E_{\mathrm{so}}=a=0$.

\subsection{Perturbative approach}\label{app:weak_perturbative}

We expand~$\mathcal R_\nu=R_0+R_{1\nu}+R_{2\nu}$ in first- and second-order corrections. Without SOC ($E_{\rm so}=0$) and with equal Fermi velocities ($a=0$), the two pairing channels decouple and inserting $R_0=\operatorname{diag}(f_1,f_2)$ in (\ref{eq:weak_riccati_G}) yields $\Delta_j f_j^2-2\epsilon f_j+\Delta_j=0$ with the solutions
\begin{equation}\label{eq:weak_scalar_quadratic}
    f_j=\frac{\epsilon-i\Omega_j}{\Delta_j}=e^{-i\alpha_j},
\end{equation}
where~$\Omega_j=\sqrt{\Delta_j^2-\epsilon^2}$ and~$\alpha_j=\arccos(\epsilon/\Delta_j)$. We collect the decay rates
in~$\Omega=\operatorname{diag}(\Omega_1,\Omega_2)$. The decaying root has a negative imaginary part. It corresponds to
the decay momenta~$q_{j\pm}=i\nu \Omega_j/(\hbar v_0)$ at the two interfaces.

At perturbative order $n\geq1$, the unknown correction~$R_{n\nu}$ enters Eq.~\eqref{eq:weak_riccati_G} through $R_{n\nu}DR_0+R_0DR_{n\nu}-2\epsilon R_{n\nu} =-i\{\Omega,R_{n\nu}\}$, while the remaining terms depend only
on lower-order corrections. The equations for the corrections therefore separate into scalar
equations for the individual matrix elements, each with the denominator~$\Omega_k+\Omega_j$. At first order this gives
\begin{equation}\label{eq:r_weak_first_order_pairing}
    (R_{1\nu})_{kj}=i\,\frac{f_j-f_k}{\Omega_k+\Omega_j}\,M_{kj}
    \bigl(\nu E_{\mathrm{so}}+i a\Omega_j\bigr),
\end{equation}
which is off-diagonal, because~$f_j-f_k$ vanishes for~$k=j$. Its SOC part changes sign between the two
interfaces, whereas the velocity part has the same sign at both interfaces. The
second-order correction follows from
\begin{equation}\label{eq:weak_second_order_sylvester}
\begin{split}
    -i&\{\Omega,R_{2\nu}\}=\nu E_{\mathrm{so}}[M,R_{1\nu}]-R_{1\nu}DR_{1\nu}\\
    &-a\epsilon\{M,R_{1\nu}\}+aR_{1\nu}MD R_0+aR_0MDR_{1\nu}.
\end{split}
\end{equation}
It contains terms proportional to~$E_{\mathrm{so}}^2$, $aE_{\mathrm{so}}$ and~$a^2$. Only the mixed term changes sign
between the interfaces. In contrast to~$R_{1\nu}$, the second-order correction also has diagonal elements.

For pure $s$-wave pairing,~$R_0$ is proportional to the identity, so that all corrections vanish for any orientation
of the axes. The same holds for parallel triplet and SOC axes, where~$[M,D]=0$. A sufficient condition for the
expansion is~$\abs{E_{\mathrm{so}}}\ll\min_j\Omega_j(\epsilon)$ together with~$\abs{a}\ll1$.

\subsection{Flux normalization and band basis}\label{app:weak_band}

For unequal Fermi velocities the matrices~$\mathcal R_\nu$ are not unitary. A wave with amplitude~$\psi$ in a channel with
velocity~$v$ carries the current~$v\abs{\psi}^2$. Below the gap the reflected current equals the incident current,
which implies~$\mathcal R_\nu^\dagger F\mathcal R_\nu=F$ rather than~$\mathcal R_\nu^\dagger\mathcal R_\nu=\mathbbm{1}$.
Measuring amplitudes in units of the square root of the current restores unitarity. With~$B=F/v_0=\mathbbm{1}+aM$ the
flux-normalized matrices are
\begin{equation}\label{eq:weak_flux_normalization}
    S_\nu=B^{1/2}\mathcal R_\nu B^{-1/2},\qquad S_\nu^\dagger S_\nu=\mathbbm{1}.
\end{equation}
Expanding the square roots gives~$S_\nu=R_0+\widetilde R_{1\nu}+\widetilde R_{2\nu}$ with
\begin{equation}\label{eq:weak_flux_corrections}
    \begin{aligned}
        \widetilde R_{1\nu}&=R_{1\nu}+\frac{a}{2}[M,R_0],\\
        \widetilde R_{2\nu}&=R_{2\nu}+\frac{a}{2}[M,R_{1\nu}]+\frac{a^2}{4}(R_0-MR_0M).
    \end{aligned}
\end{equation}
At zero field the normal-region transfer matrices commute with the band velocities. The change of normalization
therefore transforms the round trip by a similarity transformation and leaves the bound-state energies unchanged.

Restoring the lead phases and rotating to the band basis gives the reflection matrices of the main text,
\begin{equation}\label{eq:r_weak_soc_correction}
    \hat r_{\mathrm{he}}=e^{-i\varphi_{\mathrm{R}}}W^\dagger S_+W,\qquad
    \hat r_{\mathrm{eh}}=e^{i\varphi_{\mathrm{L}}}W^\dagger S_-W,
\end{equation}
where the rotation matrix $W$ is defined below Eq.~(\ref{eq:band_spinors}).

\subsection{Results}\label{app:weak_parameters}

At zeroth order these reflection matrices reduce to
\begin{align}\label{eq:r0_app}
\hat r_{\mathrm{he}}={} & {} e^{-i\varphi_{\mathrm{R}}}e^{-i\alpha_+}
    e^{-i\alpha_-\hat{\boldsymbol{d}}\cdot\boldsymbol{\sigma}},\\
    \hat r_{\mathrm{eh}}={} & {} e^{i\varphi_{\mathrm{L}}}e^{-i\alpha_+}
    e^{-i\alpha_-\hat{\boldsymbol{d}}\cdot\boldsymbol{\sigma}},
\end{align}
with~$\alpha_\pm=\tfrac12(\alpha_1\pm\alpha_2)$.
This is Eq.~\eqref{eq:r_weak0} of the main text with~$\chi=\alpha_+$, $\lambda=\alpha_-$ and~$\hat{\boldsymbol b}_\pm=\hat{\boldsymbol d}$.

Any unitary~$2\times2$ matrix is a phase times an SU(2) rotation. To second order the flux-normalized
matrices therefore keep the form of Eq.~\eqref{eq:r_weak} and the corrections enter the angles $\chi$ and $\lambda$,
\begin{align}
    \chi&=\alpha_++ \frac{\tilde \epsilon b^2 \sin^2 \theta}{\omega_1\omega_2(\omega_1+\omega_2)} \bigg( \sin^2 \alpha_- +\frac{2p^2}{1-p^2} \bigg),\label{eq:weak_corrected_chi}\\
    \lambda &= \alpha_--\frac{\tilde \epsilon b^2 \sin^2 \theta}{\omega_1\omega_2(\omega_1+\omega_2)} \frac{p}{1-p^2} \bigg[1 + \frac{4p^2}{(\omega_1+\omega_2)^2} \bigg],\label{eq:weak_corrected_beta}
\end{align}
and the vector $\hat{\boldsymbol b}_\pm$,
\begin{align}
    \hat{\boldsymbol b}_\nu&=w(-\sin\Theta\,\hat{\boldsymbol e}_x
    +\cos\Theta\,\hat{\boldsymbol e}_z)+\nu u\,\hat{\boldsymbol e}_y,\label{eq:weak_corrected_axis}
\end{align}
with
\begin{align}
    u&=\frac{2b \sin\theta}{\omega_1+\omega_2} (1+2ar \cos \theta), \quad w=\sqrt{1-u^2},\\
    \Theta&=\theta + ra\sin\theta - \frac{2b^2 \sin2\theta}{(\omega_1+\omega_2)^2}.
\end{align}
The small spin-orbit-induced parameters are $a$ and $b=E_{\rm so} / \Delta_{\rm S}$, and we neglected small relative contributions $\propto (a/b)^2 \sim (\Delta_{\rm S}/\mu)^2$.
We renormalized all energies similarly, $\tilde \epsilon = \epsilon/  \Delta_{\rm S}$, $\omega_{1,2} = \Omega_{1,2}/ \Delta_{\rm S}$ and $p = \Delta_{\rm P}/ \Delta_{\rm S}$, and we introduced the symmetry factor $r = (\Omega_1-\Omega_2)/ (\Omega_1+\Omega_2)$.
The only difference at the two interfaces, apart from the phases $\pm \varphi_{{\rm L},{\rm R}}$, is the opposite tilt of the spin rotation axis out of the $xz$-plane.
To leading order the tilt is~$u\simeq 2 b\sin\theta/(\omega_1+\omega_2)$, which describes the
precession of the spin about the SOC axis while the quasiparticle penetrates the superconductor over its decay
length. For equal gaps~$\lambda=0$ and no spin mixing occurs.

\section{Bound-state condition}\label{app:weak_abs}

We can now include the transfer matrices for the normal region and set up the bound-state determinant condition of Eq.~(\ref{eq:ABS_det}).
All matrices in this section are expressed in the band basis, and the normal-region transfer matrices then separate into
scalar phases and spin rotations about the SOC axis,
\begin{equation}\label{eq:weak_transfer_split_app}
    t_{e,h}=e^{\pm i\bar k_{\mathrm{F}}L+i\epsilon L/(\hbar\bar v)}e^{i(\delta_\epsilon\mp \Phi)\sigma_z},
\end{equation}
with the spin-precession angle $\Phi=(k_{\mathrm{F}2}-k_{\mathrm{F}1})L/2
= E_{\mathrm{so}}L/(\hbar\bar v)$ (see Sec.~\ref{sec:weak})
and the energy-dependent correction $\delta_\epsilon=\epsilon(\tau_1-\tau_2)/2\hbar=\epsilon g_v/2$.
We define $\zeta$ through
\begin{equation}\label{eq:weak_round_trip_corrected}
     \cos\zeta=\frac12\operatorname{Tr}
     {\cal W},
\end{equation}
in terms of the round trip operator
\begin{equation}
     {\cal W} = 
     e^{i(\delta_\epsilon - \Phi)\sigma_z}
     e^{-i\lambda\hat{\boldsymbol b}_-\cdot\boldsymbol\sigma}
     e^{i(\delta_\epsilon + \Phi)\sigma_z}
     e^{-i\lambda\hat{\boldsymbol b}_+\cdot\boldsymbol\sigma},
\end{equation}
which can then be written explicitly as
\begin{align}\label{eq:zeta_weak_corrected_explicit}
    \cos\zeta={}&
    2\left(\cos\lambda\cos\delta_\epsilon+w \sin\lambda \cos\Theta \sin \delta_\epsilon\right)^2\nonumber\\
    &+2\sin^2\lambda \left( u \cos \Phi + w \sin\Theta \sin\Phi \right)^2 -1.
\end{align}
The determinant
condition~\eqref{eq:ABS_det} then yields two solutions, which follow from
\begin{equation}\label{eq:ABS_weak_second_order}
    2\pi n=\varphi+\frac{2\epsilon L}{\hbar\bar v}-2\chi+\eta\zeta,\qquad \text{with }\eta=\pm1,
\end{equation}
which is Eq.~\eqref{eq:ABS_transcendental} of the main text. The normal propagation is not expanded perturbatively, so the SOC phase
accumulated in a long normal region may be large.

\subsection{Low-energy limit}\label{app:weak_low_energy}

At small energies $\epsilon \ll \Delta_{1,2}$ the interface parameters become
\begin{align}
    \chi \approx {} & {} \frac{\pi}{2} - \frac{\tilde \epsilon}{1-p^2} \left( 1 - b^2 \frac{p^2\sin^2 \theta}{1-p^2} \right),\\
    \lambda \approx {} & {} \frac{\tilde \epsilon p}{2-2p^2} \left(2-b^2\sin^2\theta
    \frac{1+p^2}{1-p^2}\right),\\
    u \approx {} & {} b\sin\theta \left( 1+ 2 ap \cos\theta \right),\\
    \Theta \approx {} & {} \theta + ap\sin\theta - \frac{1}{2}b^2 \sin 2\theta.
\end{align}
The effective gap can then be defined as
\begin{equation}
    \Delta_{\mathrm{eff}}=\left(-\frac{\partial \chi}{\partial \epsilon} \right)^{-1} = \Delta_{\rm S}\frac{(1-p^2)^2}{1-p^2-b^2p^2\sin^2\theta},
\end{equation}
which gives Eq.~\eqref{eq:Delta_eff_weak} through second order.
The velocity mismatch $a$ does not affect this slope at this order. 

In this limit the round-trip operator becomes to leading order in $\epsilon$
\begin{equation}\label{eq:weak_low_energy_round_trip}
\begin{split}
    \mathcal W=I+2i\lambda \big\{ {} & {}  (w\sin\Theta \cos\Phi - u\sin\Phi)\sigma_\perp \\
    {} & {} - w \cos\Theta \,\sigma_z \big\} + 2i\delta_\epsilon \sigma_z,
\end{split}
\end{equation}
where $\sigma_\perp=\cos\Phi\,\sigma_x+\sin\Phi\,\sigma_y$.
The velocity contribution $\delta_\epsilon$ and the longitudinal pairing contribution act along the SOC axis, whereas the transverse pairing
contribution acts along an orthogonal axis. Expanding Eq.~\eqref{eq:weak_round_trip_corrected} accordingly in small $\epsilon$ gives
\begin{equation}\label{eq:weak_low_energy_zeta}
    \cos\zeta=1-\frac{\epsilon^2}{2}
    \left[(g_v-g_\parallel)^2+g_\perp^2\right]+O(\epsilon^4),
\end{equation}
with
\begin{align}
    g_\parallel = {} & {} \frac{2}{\Delta_{\rm S}} \frac{\lambda}{\tilde \epsilon} w \cos\Theta,\\
    g_\perp = {} & {} \frac{2}{\Delta_{\rm S}} \frac{\lambda}{\tilde \epsilon}(w\sin\Theta \cos\Phi - u\sin\Phi),
\end{align}
and $g_v = (\tau_1 - \tau_2)/\hbar$.
From Eq.~(\ref{eq:weak_low_energy_zeta}) we see that at low energies $\zeta \approx \zeta_1 \epsilon$, with $\zeta_1$ as defined in Eq.~\eqref{eq:zeta1}. Keeping only the leading order in spin-orbit coupling in the interface matrices yields $g_\parallel = (2p/\Delta_{\rm S})(\cos \theta / [1-p^2])$ and $g_\perp = (2p/\Delta_{\rm S})(\sin \theta \cos\Phi / [1-p^2])$.

To find the approximate spectrum we replace $\chi$ by $\arccos(\epsilon/\Delta_{\mathrm{eff}})$, which reproduces its constant
and linear terms in~$\epsilon$. For~$n=0$ and positive energies near~$\varphi=\pi$ the bound-state condition becomes
\begin{equation}\label{eq:weak_low_energy_arccos}
    \arccos\!\left(\frac{\epsilon}{\Delta_{\mathrm{eff}}}\right)
    \simeq\frac{\varphi}{2}+\epsilon\left(\frac{L}{\hbar\bar v}+\eta\frac{\zeta_1}{2}\right).
\end{equation}
Taking the cosine and expanding the additional phase to first order gives Eq.~\eqref{eq:ABS_approx}. This formally requires
\begin{equation}
\left|\epsilon\left[ \frac{L}{\hbar\bar v} \pm \frac{1}{2}\sqrt{(g_v-g_\parallel)^2+g_\perp^2} \right] \right| \ll1,
\end{equation}
which can hold near~$\varphi=\pi$ also for long junctions.

\subsection{Zeeman field in the normal region}\label{app:weak_zeeman}

A Zeeman field confined to the normal segment leaves the interface parameters of App.~\ref{app:weak_parameters}
unchanged. We use the zero-field band basis with the velocity matrix~$v_{\mathrm{band}}=v_0(I+a\sigma_z)$. The electron
and hole blocks of the linearized normal-region Hamiltonian are
\begin{equation}\label{eq:weak_zeeman_normal_blocks}
 \begin{aligned}
    H_e(q)&=\hbar qv_{\mathrm{band}}+E_{\mathrm{so}}\sigma_z+Z,\\
    H_h(q)&=-\hbar qv_{\mathrm{band}}-E_{\mathrm{so}}\sigma_z+Z,
 \end{aligned}
\end{equation}
where~$Z=\mathbf B_{\mathrm Z}\cdot\boldsymbol\sigma$ in this basis. Including the flux normalization the
transfer matrices are
\begin{equation}\label{eq:weak_zeeman_transfer}
    t_{e,h}=\exp\!\left[\frac{iL}{\hbar}
    v_{\mathrm{band}}^{-1/2}(\epsilon I\mp E_{\mathrm{so}}\sigma_z-Z)v_{\mathrm{band}}^{-1/2}\right].
\end{equation}
The upper sign refers to the electron and the lower sign to the hole. The reference factors~$e^{\pm ik_{\mathrm F}L}$
cancel over the round trip and are omitted. The generators are Hermitian, so the transfer matrices are unitary also
for a field transverse to the SOC axis.

With~$B_\parallel=\mathbf B_{\mathrm Z}\cdot\hat{\boldsymbol m}_{\mathrm{so}}$ and
$\mathbf B_\perp=\mathbf B_{\mathrm Z}-B_\parallel\hat{\boldsymbol m}_{\mathrm{so}}$, separating the scalar phase from the
spin rotation gives
\begin{equation}\label{eq:weak_zeeman_propagation}
        t_{e,h}=e^{i\gamma_{e,h}}e^{-i\boldsymbol{\mathcal B}_\pm\cdot\boldsymbol\sigma},
\end{equation}
with
\begin{align}
        \gamma_{e,h}= {} & {} \frac{L}{\hbar\bar v}(\epsilon+aB_\parallel\pm aE_{\rm so}),\\
        \boldsymbol{\mathcal B}_{\pm}
        = {} & {} \frac{L}{\hbar\bar v} \left[ (B_\parallel\pm E_{\mathrm{so}}+a\epsilon)\hat{\boldsymbol m}_{\mathrm{so}}
        +\sqrt{1-a^2}\,\mathbf B_\perp \right].
\end{align}
For~$a=0$ the effective fields reduce to Eq.~\eqref{eq:Bfields} times~$L/(\hbar \bar v)$. For~$\mathbf B_{\mathrm Z}=0$ the rotations reproduce Eq.~\eqref{eq:weak_transfer_split_app}.
Unequal velocities also produce a field-dependent common phase,
\begin{equation}\label{eq:weak_zeeman_scalar_phase}
    \begin{aligned}
        \gamma_e+\gamma_h&=\frac{2\epsilon L}{\hbar\bar v}+\delta\varphi_B,\\
        \delta\varphi_B&=\frac{2aB_\parallel L}{\hbar\bar v}
        =\frac{B_\parallel L}{\hbar}
        \left(\frac{1}{v_{\mathrm{F}2}}-\frac{1}{v_{\mathrm{F}1}}\right).
    \end{aligned}
\end{equation}

The phase $\zeta_B$ including the effect of the Zeeman field follows again from $\cos \zeta_B = \frac{1}{2} {\rm Tr} {\cal W}_B$, where
\begin{equation}
    {\cal W}_B = e^{-i\boldsymbol{\mathcal B}_+\cdot\boldsymbol\sigma}
    e^{-i\lambda\hat{\boldsymbol b}_-\cdot\boldsymbol\sigma}
    e^{-i\boldsymbol{\mathcal B}_-\cdot\boldsymbol\sigma}
    e^{-i\lambda\hat{\boldsymbol b}_+\cdot\boldsymbol\sigma}
\end{equation}
is the round-trip operator in the presence of a Zeeman field.
Explicitly, one finds
\begin{equation}
    \cos \zeta_B = A_+A_- - {\bf A}_+ \cdot {\bf A}_-,
\end{equation}
with
\begin{align}
    A_\pm = {} & {} \cos {\cal B}_\pm \cos\lambda-\sin{\cal B}_\pm \sin\lambda\,
        \hat{\boldsymbol n}_\pm\cdot\hat{\boldsymbol b}_\mp,\\
    {\bf A}_\pm = {} & {} \sin {\cal B}_\pm \cos\lambda\,\hat{\boldsymbol n}_\pm         +\cos {\cal B}_\pm \sin\lambda\,\hat{\boldsymbol b}_\mp \nonumber\\
    {} & {}  +\sin {\cal B}_\pm \sin\lambda\, \hat{\boldsymbol n}_\pm\times\hat{\boldsymbol b}_\mp,
\end{align}
with ${\cal B}_\pm = |\boldsymbol{\mathcal{B}}_\pm|$ and $\hat{\boldsymbol n}_\pm = \boldsymbol{\mathcal{B}}_\pm / {\cal B}_\pm$ the magnitude and direction of the fields $\boldsymbol{\mathcal{B}}_\pm$.
Combined with the scalar phases this gives Eq.~\eqref{eq:ABS_weak_second_order_B_main}. At zero energy~$\lambda=0$ and the
round-trip operator reduces to~$e^{-i\boldsymbol{\mathcal B}_+\cdot\boldsymbol\sigma}e^{-i\boldsymbol{\mathcal B}_-\cdot\boldsymbol\sigma}$. This rotation vanishes at zero field but not at finite field. For a field along
the SOC axis, i.e., ${\bf B}_\perp = 0$, one finds $\cos\zeta_B(\epsilon=0)=\cos[2B_\parallel L/(\hbar\bar v)]$. The low-energy expansion $\zeta\simeq\epsilon\zeta_1$ we derived at zero field
therefore does not apply at finite field. Only keeping the leading contributions to the interface matrices, the same expressions as derived in this section hold
with~$\hat{\boldsymbol b}_\pm=\hat{\boldsymbol d}$ and~$\lambda=\alpha_-$.

\section{Tight-binding model}
\label{app:tb}

The tight-binding results in Figs.~\ref{fig:abs_energies}, \ref{fig:splitting} and~\ref{fig:splitting_soc} follow from
discretizing the BdG Hamiltonian~\eqref{eq:BdG} on a chain with lattice spacing~$\delta x$. In momentum space the
lattice model reads
\begin{equation}\label{eq:tb_model}
\begin{split}
    \xi(k) &= 2t\left[1-\cos(k\,\delta x)\right] - \mu, \\
    E_{\mathrm{so}}(k) &= t_{\mathrm{so},1}\sin(k\,\delta x) + t_{\mathrm{so},3}\sin(3k\,\delta x), \\
    \hat{\Delta}(k) &= \left[\Delta_{\mathrm{S}}\mathbbm{1}_2
    + \Delta_{\mathrm{P}}'\sin(k\,\delta x)\,\hat{\boldsymbol{d}}\cdot\boldsymbol{\sigma}\right]e^{i\varphi(x)}.
\end{split}
\end{equation}
In real space this model yields a nearest-neighbor hopping~$t=\hbar^2/(2m\,\delta x^2)$, a spin-dependent
hopping between first and third neighbors, an on-site singlet pairing and a triplet pairing on nearest-neighbor bonds.
The chemical potential~$\mu$ is measured from the bottom of the band, as in Eq.~\eqref{eq:H0_full}. The pairing is
nonzero only in the two leads, where it carries the phases~$\pm\varphi/2$.

We use the reference units~$a_0$ and~$t_0=\hbar^2/(2ma_0^2)$ and the fine spacing~$\delta x=a_0/10$, so that
$t=100\,t_0$. To leading order in~$k\delta x$, Eq.~\eqref{eq:tb_model} reproduces the continuum model with
$\alpha_{\mathrm{so}}=(t_{\mathrm{so},1}+3t_{\mathrm{so},3})\,\delta x$, $\beta_{\mathrm{so}}=-4t_{\mathrm{so},3}\,\delta x^3$
and $\Delta_{\mathrm{P}}=\Delta_{\mathrm{P}}'\,k_{\mathrm F}\delta x$. All Fermi points satisfy~$k_{\mathrm{F}j}\delta x<0.18$, so
lattice corrections are small. Each lead has a length of~$350\,a_0$ (3500 sites), several times the decay length of
the bound states. The normal region has~$L=25a_0$ (250 sites). We diagonalize the Hamiltonian numerically and
identify the Andreev levels as the subgap eigenstates with more than half of their weight within~$100\,a_0$ of the
normal region.

\begin{table}[t]
\begin{ruledtabular}
\begin{tabular}{lcc}
 & strong SOC & weak SOC \\
\hline
$E_{\mathrm{so}}/\Delta_{\mathrm S}$ & $14$ & $0.15$ \\
$\alpha_{\mathrm{so}}$ ($t_0a_0$) & $0.703$ & $7.54\times10^{-3}$ \\
$\beta_{\mathrm{so}}$ ($t_0a_0^3$) & $-5.31\times10^{-2}$ & $-5.69\times10^{-4}$ \\
$k_{\mathrm{F}1},\,k_{\mathrm{F}2}$ & $0.811,\,1.433$ & $1.093,\,1.099$ \\
$v_{\mathrm{F}1},\,v_{\mathrm{F}2}$ & $2.218,\,2.481$ & $2.186,\,2.189$ \\
$\Delta_1,\,\Delta_2$ ($\Delta_{\mathrm S}$) & $1.19,\,0.67$ & $1.33,\,0.67$ \\
$L/\xi$ & $0.53$ & $0.57$ \\
\end{tabular}
\end{ruledtabular}
\caption{Parameters of the tight-binding benchmarks. All calculations use~$\Delta_{\mathrm S}=0.05\,t_0$,
$\Delta_{\mathrm P}=0.33\,\Delta_{\mathrm S}$ ($\Delta_{\mathrm P}'=0.15\,t_0$), $\theta=0.7$, $L=25\,a_0$ and the lead
chemical potential~$\mu_{\mathrm S}=1.2\,t_0$, which gives~$k_{\mathrm F}=1.10\,a_0^{-1}$. The normal region has the lead chemical potential, so the values of~$k_{\mathrm{F}j}$ and~$v_{\mathrm{F}j}$ apply in both regions.
The gaps are the helicity gaps of Eq.~\eqref{eq:helicity_gaps} for strong and~$\Delta_{\mathrm S}\pm\Delta_{\mathrm P}$
for weak SOC. Momenta are in units of~$a_0^{-1}$ and velocities in units of~$t_0a_0/\hbar$.}
\label{tab:tb}
\end{table}

Table~\ref{tab:tb} lists the parameters. We calibrate~$E_{\mathrm{so}}$ on the lattice dispersion at the lead
momentum~$k_{\mathrm F}$ and keep the ratio~$\beta_{\mathrm{so}}/\alpha_{\mathrm{so}}=-0.075\,a_0^2$ fixed throughout.
The analytical curves take every quantity that enters the Andreev reflection from the leads. These are the gaps, with
$k_{\mathrm{F}j}/k_{\mathrm F}$ in Eq.~\eqref{eq:helicity_gaps} evaluated at the lead Fermi points, as well as the
SOC energy and the velocity asymmetry of the weak-SOC reflection matrices. The coherence length is~$\xi=\hbar\bar v_{\mathrm F}/\Delta_{\mathrm S}$ with the harmonic mean Fermi velocity.

In Fig.~\ref{fig:abs_energies}, panel~(a) uses the strong-SOC band structure, $E_{\mathrm{so}}=14\,\Delta_{\mathrm S}$, with~$\Delta_{\mathrm P}=0$.
Panel~(c) adds the Zeeman field~$B_\parallel=0.1\,\Delta_{\mathrm S}$ in the normal region. The green curves of Fig.~\ref{fig:splitting} reverse the sign of~$\Delta_{\mathrm P}$.

Figure~\ref{fig:splitting_soc} varies only~$E_{\mathrm{so}}$, by scaling~$\alpha_{\mathrm{so}}$ and~$\beta_{\mathrm{so}}$
together. The velocity difference~$\hbar\abs{\delta v_{\mathrm F}}=2\abs{\beta_{\mathrm{so}}}k_{\mathrm F}^2$ therefore grows
linearly with~$E_{\mathrm{so}}$. The helicity gaps change as well, because the two Fermi points move apart. From
$E_{\mathrm{so}}=5\,\Delta_{\mathrm S}$ to~$20\,\Delta_{\mathrm S}$, the velocity ratio~$v_{\mathrm F2}/v_{\mathrm F1}$
increases from~$1.04$ to~$1.17$ and the gaps change from~$(1.23,0.72)\,\Delta_{\mathrm S}$
to~$(1.16,0.64)\,\Delta_{\mathrm S}$. The increase of the splitting in Fig.~\ref{fig:splitting_soc}(a) stems from
the gaps alone, which would raise it by~$11\%$ at fixed velocities. The growing velocity mismatch reduces it slightly,
because the slower band carries the larger gap.

\section{Model of germanium holes}
\label{app:ge}

The Luttinger-Kohn Hamiltonian is discretized on a square lattice of~$20\times20$ sites with a spacing of~$1$~nm
using Kwant~\cite{grothKwantSoftwarePackage2014}. The wire is infinite along~$x$. We use the Luttinger
parameters~$\gamma_1=13.4$ and~$\gamma_2=\gamma_3\equiv\gamma_s=5$~\cite{winklerSpinOrbitCoupling2003,
babkinSuperconductingProximityEffect2025}. In addition to the confinement, an electric field~$E_y=6$~mV/nm
perpendicular to the wire induces the Rashba splitting. Strain enters through a constant Bir-Pikus term,
$H=H_{\rm LK}+H_{\rm BP}$, with~\cite{BirPikus1959, saraviaStrainSplitEnergyBands1969,
willatzenMethodElectronicProperties2009, abadillo-urielHoleSpinDrivingStrainInduced2023,
valvoElectricallyTuneableVariability2025}
\begin{equation}\label{eq:H_BP}
    H_{\rm BP}=\begin{pmatrix}
    P_\epsilon+Q_\epsilon & 0 & S_\epsilon & R_\epsilon\\
    0 & P_\epsilon+Q_\epsilon & R_\epsilon^* & -S_\epsilon^*\\
    S_\epsilon^* & R_\epsilon & P_\epsilon-Q_\epsilon & 0\\
    R_\epsilon^* & -S_\epsilon & 0 & P_\epsilon-Q_\epsilon
    \end{pmatrix}
\end{equation}
in the basis of Sec.~\ref{sec:ge}. The entries are
\begin{equation}
 \begin{gathered}
 P_\epsilon=-a_v\epsilon_A,\qquad Q_\epsilon=-\tfrac12 b_v\epsilon_D,\\
 R_\epsilon=\tfrac{\sqrt3}{2}\,b_v(\epsilon_{xx}-\epsilon_{yy})-id_v\epsilon_{xy},\\
 S_\epsilon=d_v(\epsilon_{xz}-i\epsilon_{yz}),
 \end{gathered}
\end{equation}
with~$\epsilon_A=2\epsilon_{xx}+\epsilon_{zz}$ and~$\epsilon_D=2\epsilon_{xx}-2\epsilon_{zz}$. We take the
deformation potentials~$b_v=2.16$~eV and~$d_v=6.06$~eV of Ge from
Ref.~\cite{abadillo-urielHoleSpinDrivingStrainInduced2023}, with signs adapted to the sign convention
of~$H_{LK}$. The term~$P_\epsilon$ only shifts all energies and is neglected. The strain components are taken
constant, with~$\epsilon_{xx}=\epsilon_{yy}=-0.0061$, $\epsilon_{zz}=0.001$, $\epsilon_{xy}=10^{-5}$
and~$\epsilon_{xz}=\epsilon_{yz}=10^{-4}$~\cite{valvoElectricallyTuneableVariability2025}.
This corresponds to compressive strain in the $xy$-plane and a weak tensile strain along~$z$.
We diagonalize the Hamiltonian on the lattice and obtain the Fermi momenta and velocities directly from the band structure.

The proximity-induced pairing matrix for the holes
is~\cite{babkinSuperconductingProximityEffect2025}
\begin{align}\label{eq:Delta_LK}
        \Delta_{\rm LK}&=
        \begin{pmatrix}
            \Delta_{\rm H}\mathbbm{1}_2 & \hat\Delta_{\rm HL}\\            
            \hat\Delta_{\rm HL}^\dagger & \Delta_{\rm L}\mathbbm{1}_2
        \end{pmatrix},
\end{align}
with
\begin{align}
        \hat\Delta_{\rm HL}&=
        \begin{pmatrix}
            \Delta_{\rm HL,a} & \Delta_{\rm HL,b}\\ 
            \Delta_{\rm HL,b}^* & -\Delta_{\rm HL,a}^*
        \end{pmatrix},
\end{align}
where the upper (lower) diagonal block in (\ref{eq:Delta_LK}) acts on the heavy-hole (light-hole) states. In terms of the parent gap~$\Delta_0$ of the superconductor the
entries are~\cite{babkinSuperconductingProximityEffect2025}
\begin{equation}
   \begin{split}
    \Delta_{\rm H} &= -t_+t_-\Delta_0,\\
    \Delta_{\rm L} &= -(t_+t_-+2t_z^2)\frac{\Delta_0}{3},\\
    \Delta_{\rm HL,a} &= \sqrt{\tfrac{2}{3}}\,t_-t_z\Delta_0,\\
    \Delta_{\rm HL,b} &= t_-^2\frac{\Delta_0}{\sqrt{3}},
   \end{split}
\end{equation}
with~$t_\pm=(t_x\pm it_y)/\sqrt2$. The tunnel coupling vector $\mathbf t=(t_x,t_y,t_z)$ characterizes the orbital content of the semiconductor states that are coupled to the $s$-type states in the superconductor. We use~$\mathbf t\propto(-1,1,3\sqrt{2})$, i.e., predominant coupling to the $p_z$-orbital, and we set the parent gap to~$\abs{\Delta_0}=0.2$~meV.

The matrix~$\Delta_{\rm LK}$ is the off-diagonal block of the BdG Hamiltonian in the time-reversed Nambu basis
\begin{equation}\label{eq:TR_basis}
    \begin{split}
        \Psi_k=\bigl(&c_{k,3/2},\,c_{k,-3/2},\,c_{k,1/2},\,c_{k,-1/2},\\
        &-c^{\dagger}_{-k,-3/2},\,c^{\dagger}_{-k,3/2},\,c^{\dagger}_{-k,-1/2},\,-c^{\dagger}_{-k,1/2}\bigr)^{T},
    \end{split}
\end{equation}
where~$c_{k,j_z}$ annihilates a hole in the state~$\ket{k,3/2,j_z}$. The lower half contains the time-reversed
partners of the upper half, so that the time-reversal operator, including its phases, is already built into the
basis. The off-diagonal block~$\Delta_{\rm LK}$ thus pairs a hole at~$k$ with its time-reversed partner at~$-k$.
Because the Luttinger-Kohn Hamiltonian is time-reversal symmetric, the hole block has the same eigenvectors as the
electron block in this basis. The effective pairing between band~$i$ and the time-reversed partner of band~$j$ is therefore the matrix element
of~$\Delta_{\rm LK}$ between the band eigenvectors, without any complex conjugation of~$\psi_j$. With the
eigenvectors~$\psi_i(k_x)$ of the two highest subbands this gives
\begin{equation}\label{eq:Delta_num}
    \Delta_{ij}^{\mathrm{num}}(k_x)=\sum_{n\in S}\psi_i^{\dagger}(n,k_x)\,\Delta_{LK}\,\psi_j(n,k_x),
\end{equation}
where~$S$ is the set of proximitized sites. In our thin sample we take all sites.

\vspace*{\baselineskip}
\bibliography{bibliography.bib}

\end{document}